\documentclass{article}
\usepackage{epsf}
\usepackage{a4wide}

\newcommand{\etal}{{et~al.~}}
\def\hMpc{\ifmmode{h^{-1}{\rm Mpc}}\else{$h^{-1}$Mpc}\fi}
\def\e{{\rm e}}
\def\d{{\rm d}}

\begin{document}

\title{Are galaxy distributions scale invariant?\\
{\large A perspective from dynamical systems theory} }

\author{J.L.~McCauley$^\star$\\
Sektion Physik\\
Universit\"{a}t M\"{u}nchen\\
Theresienstr. 37
D-80333 M\"{u}nchen\\
$\star$ Permanent Address: \\
Physics Department,\\
University of Houston,\\ 
Houston, Texas 77204}

\date{March 7, 1997}

\maketitle

%%%%%%%%%%%%%%%
\begin{abstract}
Unless there is evidence for fractal scaling with a single exponent
over distances $.1 \le r \le 100 \hMpc$ then the widely accepted
notion of scale invariance of the correlation integral for $.1 \le r
\le 10 \hMpc$ must be questioned. The attempt to extract a scaling
exponent $\nu$ from the correlation integral $n(r)$ by plotting
$\log(n(r))$ vs. $\log(r)$ is unreliable unless the underlying point
set is approximately monofractal. The extraction of a spectrum of
generalized dimensions $\nu_q$ from a plot of the correlation integral
generating function $G_n(q)$ by a similar procedure is probably an
indication that $G_n(q)$ does not scale at all. We explain these
assertions after defining the term multifractal,
mutually--inconsistent definitions having been confused together in
the cosmology literature. Part of this confusion is traced to a
misleading speculation made earlier in the dynamical systems theory
literature, while other errors follow from confusing together entirely
different definitions of ``multifractal'' from two different schools
of thought. Most important are serious errors in data analysis that
follow from taking for granted a largest term approximation that is
inevitably advertised in the literature on both fractals and dynamical
systems theory.
\end{abstract}

%%%%%%%%%%%%%%
\section{Introduction}
Knowlege of the three--dimensional distribution of matter in the
universe at $r > 150 \hMpc$ is limited. We do not know if the matter
distribution over scales $r \gg 150 \hMpc$ is homogeneous or isotropic
(background radiation, self--consistency of the standard model based
on the assumption of a {\em stable} uniform density, etc. do not
provide direct evidence about the distribution of visible matter in
the present epoch). For $r< 150 \hMpc$ the distribution of visible
matter is clearly inhomogeneous, with large voids and clustering, and
various analyses have produced results that are equivalent to claiming
scale invariance for the correlation integral, that $n(r) \approx
r^{\nu}$, with one school (\cite{davis:83}, \cite{peebles:principles})
reporting that $\nu \approx 1.23$ for $.1 < r < 10 \hMpc$, whereas the
other (\cite{coleman:fractal}, \cite{baryshev:94}) reports that $\nu
\approx 2$ for $2 < r < 150 \hMpc$. The correlation integral and
scaling exponent $\nu$ are defined in part
\ref{sect:correlation_integral}. Roughly speaking, one can think of
the scaling exponent $\nu$ as a correlation dimension, but not as
Hausdorff or box--counting dimension. We will explain why the reported
claims of scale invariance may be spurious unless the matter
distribution actually is, to a very good approximation,
monofractal. The first aim of this paper is to explain the need for a
more careful analysis of observational data than has heretofore been
performed. The second aim is to provide that analysis (part
\ref{sect:data}). The third is to explain and eliminate the confusion
over the term multifractal. Finally we will explain why a
``nonanalytic'' density does not rule out the use of differential
equations.

\begin{figure}
 \begin{center} \epsfxsize=8cm 
 \begin{minipage}{\epsfxsize}\epsffile{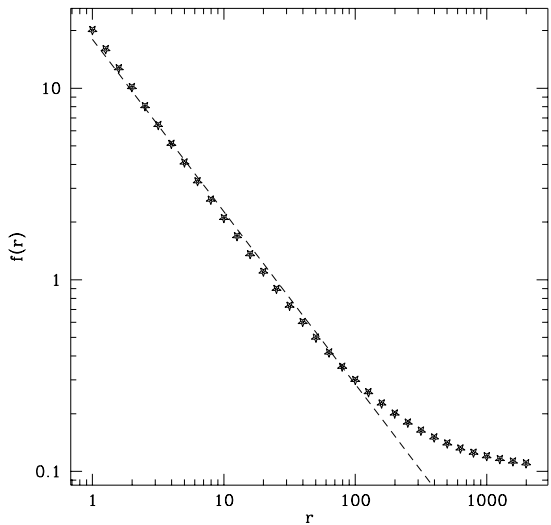}\end{minipage}
 \end{center}
\caption{ \label{fig:logplot}
Log--log plot of a function $f(r) = 20\ r^{-\gamma_1} + 0.1\
r^{-\gamma_2}$ (stars), with $\gamma_1 = 1$ and $\gamma_2 = 0$
together with the function $18\ r^{-\gamma_3}$ (dashed line) with
$\gamma_3 = 0.9$.}
\end{figure}

In any attempt to extract scaling exponents from log--log plots of
correlation or generating functions a conservative criterion in both
critical phenomena and dynamical systems theory is that linearity
should be exhibited over at least three decades, which would require
data out to at least $r = 1000~\hMpc$ in astronomy. The reason for
this is that there are too many different functions $f(r)$ that don't
scale with $r$, $f(\lambda r) \ne \lambda^\alpha f(r)$, but
$\log(f(r))$ vs. $\log(r)$ may nevertheless {\em appear} to have a
constant slope over a short enough range of $r$. The function $f(r) =
c_1 r^a + c_2 r^b$ provides a relevant example: this function is not
scale invariant because of {\em two} exponents $a$ and $b$, but it is
easy to exhibit the {\em illusion} of scale invariance by plotting
$\log(f(r))$ vs. $\log(r)$ over only two decades (see figure
\ref{fig:logplot}). If one questions the controversial claim of scale
invariance over two decades up to $r \approx 150\hMpc$, then must also
one question the {\em widely accepted} claim of scale invariance, {\em
also} over only two decades, up to $r \approx 10\hMpc$.

To make our viewpoint clear to the reader we recapitulate briefly the
controversy over scaling in this field. The earliest attempts to
extract $\nu$ were based upon the expectation of homogeneity at larger
scales, with inhomogeneities confined to $r \le r_0$ where $r_0$ is a
correlation length. By confusing an amplitude with a correlation
(i.e.~characteristic) length \cite{davis:83} $r_0 \sim 5\hMpc$ was
obtained, which is inconsistent with the observed clustering and voids
out to $r \approx 150 \hMpc$, and beyond. Coleman and Pietronero 
\cite{coleman:fractal} and Martinez \etal \cite{martinez:90-2} have
argued instead that the known data are scale invariant, and
correspondingly that there is no correlation length. Early attempts to
dismiss this point of view as the result of ``large deviations'' due
to ``unfair samples'' failed when it became clear that the apparent
scale invariance is the rule and not the exception. As a $0^{\rm th}$
order description this seems to be correct, but we show in part
\ref{sect:data} that a refinement of the method of
\cite{coleman:fractal} (see part \ref{sect:correlation_integral}) 
leads to the conclusion that the data are inadequate to draw a
conclusion for or against multifractal scaling, although it is clear
that simple fractal scaling, with a single exponent $\nu$ as claimed
by the Pietronero school, is not indicated by the data. Fractal or
not, it is clear that there is, as yet, no evidence from galaxy
redshift surveys of any crossover to homogeneity.

%%%%%%%%%%
\section{Why should anything scale?}
This is a good question because, as one expert in statistical
mechanics and nonlinear dynamics put it, if you can't calculate
anything then you can still talk about scaling. Furthermore, most
phenomena in nature do not scale, or at least have not been shown to
scale (the polls are not yet closed on the question of multifractal
scaling in the inertial (\cite{anselmet:84}; \cite{castaing:90}) and
dissipation (\cite{meneveau:87} \& \cite{meneveau:91}) ranges of fluid
turbulence, and the early returns are not entirely convincing). In
fact, there are only a few known reasons why anything should scale,
aside from dimensional analysis (Reynolds number scaling, which works
pretty well in fluid mechanics so long as one sticks to qualitative
considerations and does not look hard at the numbers). Let us
enumerate the (known) ways.

First, there is scaling of all sorts of correlation functions and
other thermodynamic quantities if you are close enough (within
$(T-T_c)/T_c$ of $10^{-3}$, at least) to a second order phase
transition. The problem with this is first that the universe is not in
thermodynamic equilibrium (tde): nonuniformities like spiral galaxies
and DNA are not generated systematically at small scales in a system
in tde. Second, there is no reason why the universe should be tuned
precisely to a critical point (where $T \approx T_c$). Critical
phenomena are popular because the scaling indices are universal,
depending only on symmetry and dimension, which allows a theorist to
forget about worse details than those that plague experimentalists and
calculate the scaling indices of real ferromagnets, for example, by
using Ising models or $\phi^4$ models (\cite{hu:82}).

We can also consider dynamical critical phenomena, where universality
classes can still be defined for scaling exponents based only on
symmetry and dimension, but that doesn't help: we are still restricted
to systems that have only very small deviations from thermal
equilibrium. Large excursions from tde aren't allowed at small scales
in these systems.

Galaxies have been modelled on the basis of critical phenomena far
from tde by using a particular cellular automaton (\cite{viscek:87})
near the percolation threshhold. Nice patterns can be produced that
look like spiral galaxies (\cite{seiden:90}), but who tunes the
galactic system to stay near the percolation threshold? This model
doesn't yet have enough physics in it to be falsifiable.

For those who believe that scaling is ubiquitous in nature, but don't
expect that Mother Nature tunes phenomena to a critical point, there
is SOC (self--organized criticality).  The idea of SOC
(\cite{bak:hownature}) is based on driven dissipative dynamical
systems far from equilibrium that quite naturally lie at a borderline
of chaos, for a large range of parameter values, and therefore require
no parameter--tuning. Criticality (a borderline of chaos) means that
all Liapunov exponents must vanish (some positive exponents are
usually allowed in the literature because models where all of them
vanish for a {\em finite} range of control parameter values
(\cite{sousa:96}) are unknown). Scaling exponents in SOC are argued
to be universal because they are expected to be
parameter--independent.  The main problem with SOC is that no one has
yet found an example of a dynamical system where the idea has been
realized, criticality without parameter tuning, criticality that
persists while parameters are varied. The SOC idea is usually
illustrated by a sandpile model that has no tunable parameters because
the parameters in the model were {\em implicitly} tuned to criticality
and then forgotten. SOC purports to provide a universal
explanation of fractal scaling indices which, if we follow the scaling
enthusiasts, should be ubiquitous in nature, but {\em fractal and
multifractal exponents are not universal and can't be used to define
universality classes}. The different models used to try to describe
the (still inadequate) data on the inertial and dissipation ranges of
fluid turbulence provide examples of this (\cite{mccauley:90} \&
\cite{mccauley:97}). Sandpile models of SOC reproduce certain
qualitative features of block spring models of earthquakes, but the
block spring models do not produce the parameter--independent
criticality demanded by SOC (\cite{crisanti:92}). SOC has not been
defined unambiguously because universality classes for SOC have not
been defined. In the absence of universality classes one cannot claim
that a simple automaton like the sandpile model represents a
complicated or complex dynamical system that occurs in nature
(\cite{mccauley:97}).

Then, there are the fractals that are generated in the phase spaces of
critical and {\em chaotic driven--dissipative} dynamical systems far
from thermal equilibrium. The scaling indices that describe the
fractals that occur in chaos are {\em not universal}. That's ok,
because while the fractal dimensions are parameter-dependent the
fractals persist (in distorted form) as the parameters are varied over
relatively wide ranges (this is usually what happens in ``SOC`` models
too). Given the coarsegrained fractal support generated by a
driven--dissipative dynamical system (``support'' of a distribution is
defined in part \ref{sect:empirical_dist}), nonuniform distributions
on that support are typically multifractal, meaning that the
coarsegrained density becomes more and more spiky (and perhaps also
intermittent with voids) as the distribution is resolved at finer and
finer scales of observation. The corresponding densities would be
nearly everywhere nondifferentiable if the mathematicians' fiction of
an infinite--precision limit were not ruled out empirically.

Conservative dynamical systems (like gravity without
dissipation/driving) cannot generate fractals in phase space: the
support of any distribution generated by a conservative dynamical
system is space--filling (Liouville's theorem). Space--filling means
that the support has the dimension of the phase space, whereas a
fractal support has a nonintegral dimension less than that of the
phase space. However, a conservative dynamical system far from thermal
equilibrium can also generate multifractal coarsegrained distributions
on the space--filling support. {\em A noninteger correlation integral
scaling exponent $\nu$ does not suggest that the galaxy distribution
has a fractal support: nonintegral $\nu$ is consistent with
multifractal distributions on space--filling supports.}

The problem with all of this is that it does not explain anything, as
yet: the fractals and multifractals discussed above all occur in the
very high dimensional phase space of all of the galaxies (each galaxy
is treated as a point particle here and below), and we do not know how
those distributions would look when projected onto the three
dimensional space of observation in astronomy. In other words, we
don't have a quantitative explanation for where fractal (including
multifractal) galaxy clusters should come from. In practice, it makes
more sense to consider hydrodynamic models of galaxy formation and
clustering. Hydrodynamics demands a coarsegrained description of the
density. Coarsegrained descriptions are precisely what are provided by
the multifractal formalism (see Vergassola \etal \cite{vergassola:94}
and references therein for a hydrodynamic approach to clustering and
voids). Unable at this time to contribute to the dynamical theory of
galaxy formation, let us forget temporarily that no theorist can yet
convincingly explain the origin of fractal galaxy clustering, if it
exists, and turn instead to the question how astronomical data should
be analyzed in order to decide the much easier question {\em whether}
fractal clustering is indicated by the observational data.

%%%%%%%%%%%
\section{Coarsegraining and fractals}

%%%%%%%%%%%
\subsection{Clustering, voids, and efficient partitions}
\label{sect:clustering_voids}

In what follows we consider only finitely many data points $N$ in some
space, the observational data. To present the ideas in the clearest
possible way we assume that the space is one--dimensional (excepting
section \ref{sect:correlation_integral} on the correlation integral,
where the dimension of space is irrelevant).  The ideas of sections
\ref{sect:clustering_voids}--\ref{sect:optimal_partitioning} are
admittedly heuristic and can only be made rigorous by the use of
generating partitions found in the phase space of certain
nonintegrable dynamical systems (\cite{cvitanovic:88}). In particular
the heuristic description is limited to one dimension (generating
partitions are not so limited), but our one dimensional treatment is
adequate to the purpose of resolving the prevailing confusion over
``multifractals'' and ``nonanalytic'' densities.

We assume that the $N$ data points in our one dimensional space are
confined to the unit interval, because any finite interval can be
converted into the unit interval by rescaling. With lengths denoted by
$l$, $0<l<1$ in all that follows (planar and three dimensional
cosmological data are also assumed to be rescaled so that $0<r<1$ in
the discussion of part \ref{sect:correlation_integral}).

Coarsegraining of the data set requires only that we cover the $N$
points by $N_n$ nonoverlapping intervals of size $l^{(n)}$. Clearly,
we can take $N_n = 2^n$ intervals of size $l^{(n)} = 2^{-n}$, $N_n =
10^n$ intervals of size $l^{(n)}=10^{-n}$, and so on. All that is
required to avoid overlap is $l^{(1)} \le 1/2$; coarsegraining per se
is not unique. The only other requirement, so far, is that we must
choose the intervals so that $l^{(n)} \ge l_{\rm min}$ where $l_{\rm
min}$ is the smallest distance between two points in the sample (a
single point cannot be coarsegrained, so that an interval containing
only one point is meaningless). We shall see that the desire of the
theorist to approximate $l_{\rm min}$ by $\epsilon$, where $\epsilon$
goes to zero, has led to serious errors in data analysis. In our
analysis $l_{\rm min}$ is always finite because it contains at least
two {\em real} data points.

In any efficient coarsegraining the intervals that cover the points
must be {\em nonoverlapping}, Coarsegrainings that are space-filling
(meaning that $N_n l^{(n)} = 1$) satisfy this criterion but do not
separate voids from clusters. For data with voids we should construct
a coarsegraining that is not space-filling, one where $N_1 l^{(1)} <
1$, in order to cover all of the clusters while excluding the largest
voids. Such a coarsegraining is more efficient than an arbitrary
one. With the desire for efficiency in mind the idea is first to
remove all of the largest voids. Then we choose the interval size
$l^{(1)} \approx (1- \sum v_{1,i} )/N_1$, where the intervals
$v_{1,i}$, all of roughly the same characteristic size, represent the
$M_1$ largest voids in the sample, and $N_1=M_1+1$. The $N_1$ first
generation intervals required to cover the $N$ data points are not
space-filling because $N_1 l^{(1)} < 1$, by construction. If, beneath
the $N_1$ intervals now covering the data, there are still voids and
clustering then we can continue systematically by removing all voids
of next largest characteristic size: in the second generation of
coarsegraining simply choose $N_2$ intervals of size $l^{(2)} \approx
(1- \sum v_{2,i})/N_2$ where the intervals $v_{2,i}$ represent the
sizes of $M_2$ largest voids covered by the $N_1$ intervals
$l^{(1)}$. This iterative procedure may be continued so long as we can
distinguish clusters from voids. There are two ways that it can
terminate. Either we reach a scale $l^{(n)} > l_{\rm min}$ where the
points are relatively evenly spaced over those intervals (so that
there is no longer a distinction between clusters and voids), or else
clustering continues all the way down to the finite limit
$l^{(n)}=l_{\rm min}$, where $l_{\rm min}$ roughly characterizes an
interparticle spacing and will be defined more precisely in part
\ref{sect:optimal_partitioning}.

The procedure outlined above describes the idea of a more efficient
partitioning than a space--filling one, a more efficient
coarsegraining of the data set because the largest voids are
systematically excluded, generation by generation in n. We do not have
only one partition but a hierarchy $n=1,2,\dots,n_{\rm max}$ of
partitions, each with interval sizes $l^{(n)}$.

\begin{figure}
 \begin{center} \epsfxsize=8cm 
 \begin{minipage}{\epsfxsize}\epsffile{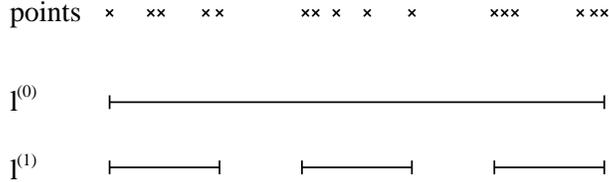}\end{minipage}
 \end{center}
\caption{ \label{fig:nonuniform}
Nonuniform clustering with uniform partitionings. Here is no way to
construct an efficient, uniform coarsgraining $l^{(2)}$.}
\end{figure}
A pencil and paper sketch of about sixteen points with two big
voids of roughly the same size, but with three or more very nonuniform
first generation clusters, e.g. figure \ref{fig:nonuniform}, shows
that the method described above will produce an efficient partition
only if the clustering is relatively uniform, only if all clusters in
a each generation are all of about the same size. When this is not the
case, when the clustering is very nonuniform (as in figure
\ref{fig:nonuniform}), then the procedure outlined above will not
produce an efficient partition and may even fail to cover the set. In
that case we could try to repair the misfit by taking $l^{(n)}$ to be
the largest of the intervals in the $n$th generation, but this may produce
overlapping intervals, $N_n l^{(n)} > 1$, which is intolerable.

We explain in the next section how``convergence problems'' arise in
the analysis of empirical data whenever arbitrary (rather than
efficient) partitions are used.

The motivation for the expectation that an {\em optimal} partition may
exist, at least in {\em some} cases, is as follows: certain
nonintegrable dynamical systems coarsegrain phase space uniquely
(\cite{feigenbaum:88}, \cite{cvitanovic:88}).  In those systems the
optimal partition is generated by the dynamics and is called the
``generating partition''. As a simple example, the invariant set of
the ternary tent map is the middle thirds Cantor set
(\cite{mccauley:chaos}). The generating partition of the ternary tent
map (obtained by $n$ backward iterations of the unit interval using
that map) is given by $N_n = 2^n$ intervals $l^{(n)} = 3^{-n}$, for
$n=1,2,\dots$. It is impossible to construct a more efficient
coarsegraining of the middle thirds Cantor set than this one. The
voids are the excluded open intervals $(1/3,2/3), (1/9,2/9),
(7/9,8/9)$, and so on (initial conditions of the ternary tent map that
lie in the voids iterate to minus infinity).

%%%%%%%
\subsection{Scale invariant clustering}
\label{sect:scale_invariant}

We restrict our considerations in this section to relatively uniform
clustering (the clusters in a generation $n$ of coarsegraining are all
of about the same size $l^{(n)}$, and the $n$th generation voids are
also all of about the same size $v_n$). Invariant quantities (scaling
exponents) can only be constructed, if at all, from $N_n$ and
$l^{(n)}$ as the generation $n$ of coarsegraining is increased, as we
look at the data set with finer and finer, but {\em never} with {\em
pointwise} resolution (the smallest interval size $l_{\rm min}$ always
contains at least {\em two} data points).

In what follows it is conceptually useful to think of the hierarchy of
intervals for $n=1,2,\dots,n_{\rm max}$ as sitting on the branches of
a tree of some order $t$. There are $N_1$ branches in generation 1,
$N_2$ in generation 2, and so on, and the tree need not be complete.
Scale invariance will be seen below to require that $N_n$ increases
exponentially, $N_n \approx t^n$, as $l^{(n)}$ decreases
exponentially, $l^{(n)} \approx a^{-n}$. Complete trees have $N_n=t^n$
branches in generation $n$ with $t \ge 2$ an integer, while incomplete
ones have noninteger $t$, in which case the order of the tree is the
next integer larger than $t$. The middle thirds Cantor set, an
idealized model of uniform clustering exhibiting big voids on all
scales, defines a complete binary tree ($t=2$).

We are only treating coarsegrained versions of $N$ points, so the
simplest kind of scale invariance is geometric (and so is the more
complicated kind, as we shall see in parts \ref{sect:multi_optimal}
and \ref{sect:multi_empirical} below).  A fine-grained picture of the
subset of any branch for $n \ge 2$ looks, upon magnification, like the
entire tree. This kind of scale invariance (\cite{mandelbrot:fractal})
is expressed by the exponent $D_0$ in
\begin{equation}
\label{eq:1}
N_n \left( l^{(n)} \right)^{D_0} \approx 1 .
\end{equation}
In other words, $D_0 \approx \ln(t)/\ln(a)$ is an exponent that
reflects self--similarity of a hierarchy of relatively {\em uniform}
clusters (clusters within clusters within \dots ).

In order to check whether (\ref{eq:1}) holds approximately for a given
(very uniform) data set it is very useful first to find as
efficient a partition as possible. The practical difference
(emphasizing the famous ``convergence'' problem) between the use of
efficient and inefficient partitions is best illustrated by an
idealized example.

\begin{figure}
\begin{center} \epsfxsize=8cm 
\begin{minipage}{\epsfxsize}\epsffile{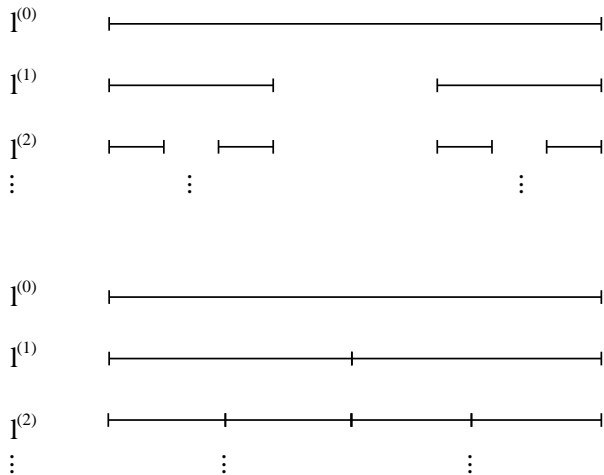}\end{minipage}
\end{center}
\caption{ \label{fig:optimal_partitioning}
On top we show the optimal partitioning for the middle--thirds Cantor
set, on the bottom a very inefficient uniform partitioning.}
\end{figure}

As an example of an optimal partition consider a very artificial data
set constructed as follows: arbitrarily choose a finite number $N$ of
points generated by ternary expansions of the form $x = .\epsilon_1
\dots \epsilon_N \dots$ with $\epsilon_i = 0$ or 2 ($\epsilon_i = 1$ is
excluded). These numbers belong to the middle--thirds Cantor set (all
ternary numbers of this kind define the middle--thirds Cantor set
\cite{gelbaum:counterexamples}).  Terminating strings $\epsilon_1 \dots 
\epsilon_N 00000 \dots$ define $N_n = 2^n$ intervals $l^{(n)} = 3^{-n}$ 
given by $[0,1/3]$ and $[2/3,1]$ in generation one ($n=1$), $[0,1/9]$,
$[2/9,1/3]$, $[2/3,7/9]$, and $[8/9,1]$ for $n = 2$, and so
on. Rational numbers of the form $x = .\overline{\epsilon_1 \epsilon_2
\dots \epsilon_N}$ (periodic expansions like $x = .020202\dots$ ) and
irrational ones $x = .\epsilon_1 \dots \epsilon_N \dots$ (where the
digit string is nonperiodic) are also covered by all of the $N_n$
intervals as well, so that the covering provided by those $N_n$
intervals $l^{(n)} = 3^{-n}$ is optimal, generation by generation.
The scaling law (\ref{eq:1}) yields $D_0 = \ln2/\ln3$. This example
illustrates a ``monofractal'' because the optimal covering is uniform
(all $N_n$ of the optimal intervals in one generation have the same
size $l^{(n)}$).

In contrast, we can try to estimate $D_0$ by using the uniform
nonoptimal covering $l^{(n)} = 2^{-n}$, a space-filling partition
given by $[0,1/2]$ and $[1/2,1]$ for $n=1$, $[0,1/4], [1/4,1/2],
[1/2,3/4]$, and $[3/4,1]$ for $n=2$, etc., that ignores the voids
alltogether. Here, with $N$ points in the data set $N \gg N_n \gg 1$
must be very large before we can expect to observe scaling with an
exponent close to $D_0=\ln2/\ln3$: for $N = 16$ points, e.g., and
using $N_n \approx l^{-D}$ then from $n = 1$ and 2 one gets $D = 1$,
$n =3$ yields $D = .86$, and further attempts to extract $D_0$ are
impossible unless the number $N$ of data points is increased. This
illustrates why, in practice only relatively efficient partitions are
of interest. We return next to the search for the optimal partition of
a typically nonuniform empirical data set of $N$ points.

%%%%%%%%%%
\subsection{The optimal partition of an empirical data set}
\label{sect:optimal_partitioning}

\begin{figure}
 \begin{center} \epsfxsize=8cm 
 \begin{minipage}{\epsfxsize}\epsffile{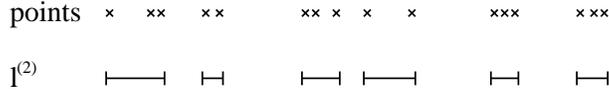}\end{minipage}
 \end{center}
\caption{ \label{fig:nonuniform2}
An efficient (nonuniform) partition $l^{(2)}$ for the point set shown
in Figure \ref{fig:nonuniform}.}
\end{figure}

For a typical empirical data set of $N$ points the most efficient
partition that one can construct will rarely be uniform. Let $l^{(n)}$
denote the size of the largest cluster in generation $n$ after
deleting the largest voids, as in part
\ref{sect:clustering_voids}. This may yield an overlapping partition
of $N_n$ uniform intervals $l^{(n)}$ where $N_n l^{(n)} > 1$, but we
can immediately improve upon that coarsegraining: in any generation
$n$ the $N_n$ intervals so--constructed will (excepting the largest
interval, which determines $l^{(n)}$) not end on data points but will
extend beyond them. To make the covering efficient simply shrink each
interval until it ends on the nearest two points of the cluster that
it was intended to cover in the first place. The result is that the
number of intervals is exactly the same as before, but we now have
$N_n$ nonuniform intervals obeying both $l_1 + \dots + l_{N_n} < 1$
and $l_1 + \dots l_{N_n} < N_n l^{(n)}$.  In other words, we have
minimized the sum $l_1+ \dots + l_{N_n}$ while holding $N_n$ fixed. It
is hard to see how a more efficient covering can be found, so for the
purpose of this paper we call the hierarchy of intervals,
so-constructed, the optimal partition (see Figure \ref{fig:nonuniform2}).

Our definition of optimal partition is a finite precision realization
of an ``optimal'' $\delta$--covering (a $\delta$--covering
approximates the ``infimum''), as is used in the mathematicians'
definition of ``Hausdorff measure'' (\cite{falconer:fractal}). The
method in the cosmology literature that may come nearest to ours, in
spirit, is the minimal spanning tree method (\cite{weygart:90}).
In dynamical systems theory the optimal partition is called the
generating partition and provides the geometric or finite precision,
definition of a fractal.

Here's an idealized example of a dynamical system with a nonuniform
optimal partition (\cite{mccauley:chaos}). The generating partition of
the asymmetric tent map with slope magnitudes $a$ and $b$ is given by
$N_n=2^n$ intervals $l_m=a^{-m} b^{-(n-m)}$ where $m=0,1,2,\dots,n$,
and describes the two--scale Cantor set (there are two first
generation scales $l_1=a^{-1}$ and $l_2=b^{-1}$). The idealized data
set consists of interval end points and also of limits of infinite
sequences of interval end points (the latter corresponding to
infinitely--many backward iterations of the unit interval by the
asymmetric tent map).  This description of the asymmetric tent map is
correct if $a^{-1} + b^{-1} \le 1$ ($a^{-1} + b^{-1} =1$ means
space--filling, while $a^{-1} + b^{-1} < 1$ produces voids and
clustering, representing an idealized nonuniform ``Cantor dust''
(\cite{mandelbrot:fractal}). One object of \cite{cvitanovic:88} was to
extract the generating partition of the Henon map.

The generalization of the scaling law (\ref{eq:1}) to our hierarchy of
nonuniform optimal partitions $\{ l_i \}$ is
\begin{equation}
\label{eq:1b}
\sum_{i=1}^{N_N} l_i^{D_H} \approx 1 ,
\end{equation}
where the scaling index $D_H$ is called the Hausdorff dimension
(\cite{falconer:fractal}). Whether a data set has a Hausdorff
dimension can only be answered empirically, by constructing the
optimal partition and checking to see whether (\ref{eq:1b}) holds over
many different generations n with the same exponent $D_H$.

The Hausdorff dimension of the two--scale Cantor set is given exactly
in the first generation by $a^{-D_H} + b^{-D_H} = 1$. In the empirical
case, in contrast, $N \gg N_n \gg 1$ is usually required in order that
we have enough data points to see scaling via either (\ref{eq:1}) or
(\ref{eq:1b}) which, as Mandelbrot \cite{mandelbrot:fractal} pointed
out early in his book, is usually confined to an intermediate range of
interval sizes $l^{(n)}$ where $l_{\rm max} \gg l^{(n)} \gg l_{\rm
min}$. Here, $l_{\rm max}$ is on the order of the size of the sample,
$l^{(n)}$ is the largest of the $N_n$ $n$th generation intervals $\{
l_i \}$, and $l_{\rm min}$ is in our case the smallest distance
between two points in the sample. In general, we should not expect to
observe scaling unless the largest intervals are much smaller than the
size of the system, and unless all of the smallest ones contain more
than a single data point. However, the larger limit may not be
applicable to present astronomical data because the systems of
galactic clusters and voids are presumably much larger than any
available sample size. Also, there is nothing to prevent our checking
for scaling all the way down to $l_{\rm min}$.  Again, if an
inefficient partition is chosen then scaling may not be observed even
if the data set is fractal, because the number $n$ of generations
needed for ``convergence'' to a scaling exponent $D_0$ or $D_H$ may
exceed the number $N$ of data points in the sample.

{\em All definitions and approximations based upon the $l^{(n)}
\rightarrow 0$ limit are systematically avoided because, as we explain
below, they lead to formulae that generally do not apply to finite
data sets.}

Suppose that we have found the optimal nonuniform partition for 
a data set. If we replace the uneven intervals in a nonuniform 
partition $\{ l_i \}$ by the largest scale of each generation 
(simply call it $l^{(n)}$), then we obtain a less optimal uniform 
covering defined by
\begin{equation}
\label{eq:old1}
N_n \left( l^{(n)} \right)^{D_0} \approx 1 .
\end{equation}
Because $N_n$ is the same as in (\ref{eq:1b}), and because $l^{(n)}
\ge l_i$, it follows that $D_0 \ge D_H$ where $D_0$ is called the
box--counting dimension. This procedure defines the most efficient
{\em uniform} partitioning of the data set if the clustering is
relatively uniform. In other words, whether or not $D_0$ provides a
good estimate of $D_H$ for low values of $n \ll N$ depends on whether
or not the uniform partition closely approximates the nonuniform
optimal one. This replacement amounts to the pointwise approximation
of a nonuniform fractal data set by a monofractal. The resulting
idealized data set can be thought of as consisting of the end points
of the uniform intervals $l^{(n)}$. For the two-scale Cantor set with
$a<b$ the box-counting dimension is $D_0=\ln 2/\ln a$. In the
two-scale Cantor set $D_0 = \ln 2/ \ln a$ and $a^{-D_H} + b^{-D_H} =1$
yield $D_0 \approx D_H$ only if $a$ and $b$ are approximately equal.

How many generations are necessary in order to convince hardened
sceptics that scaling has been observed? In both critical phenomena
and dynamical systems theory the rule of thumb is three decades on a
log--log plot, requiring astronomical data, e.g., from .1 to
100~\hMpc, or from 1 to 1000~\hMpc\ depending on the smallest distance
reported in a given catalog of galaxies. With $l^{(n)} \approx
a^{-n}$, $a \ge 2$, we would need $n \approx 3 \ln 10/\ln a$
generations.  We call this criterion ``the Geilo Criterion'' because
it was suggested at a Geilo NATO-ASI. The Geilo criterion
is not a matter of taste: it is advocated in order to deflect
erroneous reports of scaling like that indicated in figure
\ref{fig:logplot} over only two decades, $1 \le r \le 100$.

Thinking of the optimal partition of a fractal as organized onto a
tree of some order, if we write $N_n = t^n$ then the order of the tree
is the next integer greater than or equal to $t$. If $t$ is
nonintegral then the tree is incomplete. If $l^{(n)} = a^{-n}$ then $t
= a^{D_0}$. With $a \ge 2$ the $\beta$--model of fluid turbulence lies
on an incomplete tree that is at least octal (\cite{mccauley:90}). See
\cite{castagnoli:91} for a discussion of the $\beta$--model in
cosmology.

%%%%
\subsection{Multifractal scaling (via a nonuniform optimal partition)}
\label{sect:multi_optimal}

A data set that obeys scaling (\ref{eq:1b}) with a nonuniform
optimal partition may be ``multifractal''. Multifractal is always
defined here to mean a spectrum of fractal dimensions
(\cite{halsey:86}). Each dimension in a multifractal spectrum
describes the scaling of a subset of the optimal partition (a
nonuniform fractal is decomposed disjointly into a union of other
fractals).

Given the optimal partition a multifractal spectrum $D(\lambda)$ may
be defined by parameterizing the hierarchy of coarsegrained intervals
(the parameter here is $\lambda$) so that the partition is organized
into (nonoverlapping but) interwoven sub--partitions
(\cite{artuso:89}). Each fractal dimension in the multifractal
spectrum is the Hausdorff dimension of one subset of the partition (a
multifractal is always the union of a complete set of nonoverlapping
but interwoven fractals labeled systematically by some index). To
obtain scale invariance the interval sizes $l_n$ must contract
exponentially as $n$ increases. Both the $n$th generation intervals
and their contraction rates are generally nonuniform: as an
oversimplified example let $l_i = a_i^{-n}$ denote the $i$th of $N_n$
intervals in generation $n$. This defines a simple nonuniform Cantor
set based on $N_1$ different first generation scales $l_i = a_i^{-1} =
\e^{-\lambda_i}$ if $l_1 + \dots +l_{N_1} < 1$ (in a chaotic dynamical
system the contraction rate $l_i^{(n)} \approx \e^{-n
\lambda_i}$ describes the intervals of the generating partition only
asymptotically and approximately for $n \gg 1$, representing the
inverse butterfly effect for integration backward in time along
unstable manifolds (\cite{mccauley:chaos})).

Suppose that there are $N(\lambda)$ intervals with the same
contraction exponent $\lambda$ (in a dynamical system $\lambda_i$ is
the Liapunov exponent for forward evolution in time, starting from a
specific class of initial conditions, namely, all initial conditions
that yield the same Liapunov exponent $\lambda_i$). Then $N(\lambda) =
l(\lambda)^{-D(\lambda)}$ (where $l(\lambda) = \e^{-n\lambda}$)
defines the Hausdorff dimension $D(\lambda)$ of the subset of the
partition labeled by $\lambda$. We can generalize (\ref{eq:1b}) by
writing down the generating function (\cite{artuso:89})
\begin{equation}
\label{eq:2}
Z_n(\beta) = \sum_{i=1}^{N_n} l_i^{\beta} = 
\sum_\lambda N(\lambda) l(\lambda)^\beta \approx 
\sum_\lambda \e^{s(\lambda) - \beta \lambda} 
\end{equation}
with $N(l) \approx \e^{ns(\lambda)}$ for large enough $n$, and where
$Z_n(D_H) \approx 1$ defines the Hausdorff dimension of the entire
fractal (by (\ref{eq:1b})). Note that $D_H > D(\lambda)$ because
$N(\lambda) < N_n$. This simple fractal, seen as multifractal, is the
union of a complete set of interwoven, nonoverlapping monofractals
(neighboring branches on each generation of the tree generally are
labeled by different indices $\lambda$). In the ovesimplified model
above we have $Z_n(\beta) = \left( \sum a_i^{-\beta} \right)^n$.

As an example (\cite{mccauley:chaos}), consider the two scale Cantor
set (with $N_1=2$ first generation intervals $l_1 = a^{-1} > l_2 =
b^{-1}$). In generation $n$ the optimal covering is given by the $N_n
= 2^n$ intervals with sizes $l_1^m l_2^{n-m}$, of which \mbox{$N_m =
n!/m!(n-m)!$} have the same size $l_m = l_1^m l_2^{n-m}$. Here,
$\lambda = x \ln a + (1-x) \ln b$ with $x = m/n = 0, 1/n, \dots,
1$. In each generation $n$ there are $n+1$ points in the multifractal
spectrum, not more, and the number of points grows only linearly with
$n$ (still, this eliminates bi-fractals, tri-fractals, \dots, and
$N$--fractals from our definition of ``multifractal''). Using
Stirling's approximation, so that, with $x = m/n$, $N(\lambda) \approx
\e^{ns(\lambda)}$ where
\begin{equation}
s(\lambda) \approx -x \ln x - (1-x) \ln (1-x)
\end{equation}
is the Boltzmann entropy (divided by $n$) of all intervals with the
same contraction exponent $\lambda$, we have $D(\lambda) =
s(\lambda)/\lambda$, which shows the connection of fractal dimension
to entropy ($s(\lambda) = \ln 2$ and $\lambda = \ln 3$ for the middle
thirds Cantor set). Since $t(\lambda) = \e^{s(\lambda)}$ the tree is
generally binary but incomplete for any monofractal subset labeled by
the contraction index $\lambda$ in the two-scale Cantor set.

Note that ``multifractal'' is consistent with $D_H = 1$: the support
may be space-filling, while subsets of the support are fractal ($0 <
D(\lambda) <1$). In the two-scale Cantor set $D_H = 1$ corresponds to
the space-filling condition $a^{-1} + b^{-1} = 1$.

Ideas based on the generating function (\ref{eq:2}) have been used to
analyze experiments on the transition to chaos in fluid dynamics
(\cite{glazier:88}).

The notion that multifractal scaling can be verified for small data
sets (\cite{martinez:clustering}) is a misconception: fewer data
points $N$, with less precision, {\em cannot} be required for the
determination of a {\em spectrum} of fractal dimensions than are
required for determining {\em one} dimension, say $D_H$. The error in
the claim follows from confusion and errors made in defining
``multifractal'' (see parts \ref{sect:correlation_integral},
\ref{sect:generating_functions} and \ref{sect:lognormal} below).

The term multifractal has occasionally been incorrectly defined in the
cosmology literature where at least three entirely different
generating functions are confused together (see parts
\ref{sect:correlation_integral} and \ref{sect:generating_functions})
as if their corresponding scaling exponents (whenever scaling exists)
would define universality classes independently of probability
distributions and partitionings used to define those exponents. In
\cite{coleman:fractal} a far to restrictive idea of multifractal is
presented in part 6.  Multifractal spectra were introduced in analogy
with critical exponents (\cite{halsey:86}), but the expectation of
universality (\cite{kadanoff:90}) was not realized (a restricted and
still unproven universality is merely postulated for vortex cascades
in fluid turbulence (\cite{frish:turbulence})). In the theory of
chaotic dynamical systems two examples of topologic invariants are the
tree order $t$ and its degree of incompleteness (\cite{gunaratne:90},
\cite{mccauley:chaos}).

We always restrict our formulation of the requirement for fractal and
multifractal scaling to finite $l^{(n)}$ and to finitely many data
points $N$, completely avoiding mathematically idealized results that
would require for their applicability the empirically and
computationally unattainable limit where $l^{(n)} \rightarrow 0$. We
shall see in part \ref{sect:multi_empirical} that this rules out
largest term approximations, whose validity would require values of
$l^{(n)}$ that are too much small ($l^{(n)} \ll 1$, with the range of
$l^{(n)}$ extending over at least three decades) to be consistent with
the analysis of galaxy distributions.

%%%%%%%%%%
\subsection{The empirical distribution}
\label{sect:empirical_dist}

For an observational data set there is only one pointwise probability 
distribution, the empirical distribution $P(x)$ defined by the 
$N$ data points: $P(x)$ is simply the fraction of points lying to 
the left of (and including) $x$, so that $P(0)=1/N$ and $P(1)=1$, by 
construction. The distribution is constant on the voids and increases 
discontinuously at each data point, so that the plot of $P(x)$ 
is a staircase of $N-2$ steps of finite width. The data staircase 
has the singular pointwise density $\rho(x) = P'(x)$ given by
\begin{equation}
\label{eq:3}
\d P(x) = \frac{\d x}{N} \sum_{i=1}^{N} \delta(x-x_i)
\end{equation}
Of course, (\ref{eq:3}) is only a theorist's fiction: it neglects the
error bars in the locations of positions. In reality each position is
specified empirically by a finite {\em interval} whose width is the
uncertainty in location. We assume here that these uncertainties are
very small relative to the smallest separation $l_{\rm min}$ between
data points. Otherwise coarsegraining and fractal/multifractal
analysis are impossible. The density (\ref{eq:3}) will be used to
correct a more serious theoretical error in part
\ref{sect:generating_functions} below.

All that we need in what follows is the staircase $P(x)$ along with
an empirical technique for characterizing voids and clusters.  We
emphasize that attempts to ``smooth'' the staircase (via ``splines'',
e,g) will discard important information about clustering. {\em No pointwise
probability distribution other than the staircase $P(x)$},
\begin{equation}
\label{eq:3b}
P(x) = \frac{1}{N} \sum_{i=1}^{N} \Theta(x-x_i)
\end{equation}
{\em is relevant for empirical data analysis.} 

An arbitrary distribution $P(x)$ of $N$ data points is generally not
approximable in either the continuum (infinite precision) or
hydrodynamic (coarsegrained) limit by a differentiable distribution.
We shall find next that the coarsegrained versions of the empirical
distribution $P(x)$ are typically too spiky (too intermittent) to be
approximable by an everywhere differentiable distribution (reminding
us more of ``noise'' than of analyticity), even if $D_H$ is an integer
(even if the optimal partition is space--filling). The
spikiness/intermittence represent clustering and voids in the
sample. Hydrodynamics demands a coarsegrained description of a
pointwise distribution, and we will discuss in part
\ref{sect:homogeneity} how the required densities can be defined.

On an optimal or at least efficient partition a coarsegrained
probability $P_i$ is defined by the difference $P_i=\Delta P(x)$ over
the $i$th closed interval of size $l^{(n)}$, and is just the fraction
$P_i=n_i/N$ of the total number of data points $n_i$ in the $i$th
interval including the end points. While each empirical distribution
$P(x)$ is a staircase of finitely-many steps, each coarsegrained
distribution $\{ P_i \}$ is a histogram on a finite support.

The optimal partition optimally defines the ``support'' of the
hierarchy of coarsegrained empirical distributions $\{ P_i \}$: for
the optimal partition, each interval end point coincides with a point
where the staircase function $P(x)$ increases discontinuously.
Whether an empirically-constructed optimal partition is the generating
partition of a deterministic dynamical system is a separate question
(the main question, but very hard to answer (\cite{cvitanovic:88},
\cite{glazier:88})).

The coarsegrained probabilities $\{ P_i \}$ can be used to perform
averages that ignore the details of the dynamics at all scales smaller
than $l^{(n)}$ (coarsegraining the smaller scales is required in order
to define hydrodynamics). The only limitation, so far, is that $1/2
\ge l^{(n)} \ge l_{\rm min}$. Bear in mind however, that before
reaching the smallest scale $l_{\rm min}$, as $n$ is increased, we may
not be able to distinguish clustering from voids without ambiguity.
Even if clustering and voids are present at all scales they may not be
scale invariant. The construction of efficient and even optimal
partitions does not presume scale invariance, rather, the converse is
true, especially in practice.

We have used the frequency definition of probabilities because 
it arises naturally in both empirical data analysis and computer 
simulations. Using our simple example above, however, we can 
offer as an idealized staircase distribution the Cantor function 
(\cite{gelbaum:counterexamples}, \cite{mccauley:chaos}) 
\begin{equation}
\label{eq:4}
P(x) = . \frac{\epsilon_1}{2} \frac{\epsilon_2}{2} \dots 
 \frac{\epsilon_N}{2} \dots ,
\end{equation}
where, because $x = .\epsilon_1 \epsilon_2 \dots \epsilon_N \dots$ is
a ternary number with $\epsilon_i = 0$ or 2, $P(x)$ is a binary number
(because $\epsilon_i/2 = 0$ or 1). This staircase describes a
mathematician's idealization of empirical data, namely, one (of
infinitely--many) distribution that can be constructed by using points
in the middle--thirds Cantor set: $P(0)=0$, $P(1) =1$, $P(x)$ is
constant on the open voids and increases discontinuously at the end
point of any closed interval $l^{(n)} = 3^{-n}$ of the optimal
covering, where the change in $P(x)$ is $\Delta P =
2^{-n}$. Therefore, the Cantor function defines a hierarchy of {\em
uniform} coarsegrained distributions $P_i = \Delta P(x)= N_n^{-1} =
2^{-n}$, generation by generation in $n$, on the fractal support
$l^{(n)} = 3^{-n}$. The plot of the Cantor function, the
mathematicians' idealization of an empirical distribution, is called a
{\em devil's} staircase because it has $2^\infty$ steps.

\begin{figure}
\begin{center} \epsfxsize=8cm 
\begin{minipage}{\epsfxsize}\epsffile{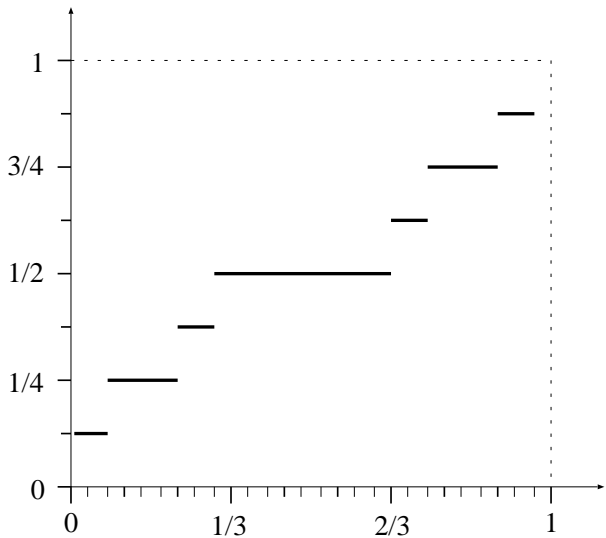}\end{minipage}
\end{center}
\caption{ \label{fig:staircase}
The idealized empirical distribution according to eq.~\ref{eq:4} with $N=3$.}
\end{figure}

A good way to look for voids is simply to plot the empirical
staircase: $P(x)$ is constant on the voids, where there are no data
points. This is illustrated by the idealized example described above
(see also figure \ref{fig:staircase}). Unfortunately this method is
not trivially generalizable to more than one dimensions.

The Cantor distribution reflects statistical independence based upon
the two first generation probabilities $p_1 = p_2 = 1/2$ for occupying
the two first generation intervals, each of length $l^{(1)} = 1/3$,
and is generated by the ternary tent map for a special class of
initial conditions. For other initial conditions the map yields other
distributions. Nonuniform distributions that lack statistical
independence are trivially easy to construct via either the ternary
Bernoulli shift or the ternary tent map. For example, iterate the
(ternary) initial condition $x_0 =
.202002000200002\dots$. Statistically independent distributions where
$p_1 > p_2$ can also be constructed with only a bit more effort (see
ch. 9 of \cite{mccauley:chaos} for the method).

Summarizing, in the beginning there is only a collection of $N$ points
(or a time series) generated by some generally unknown dynamical
system. We can construct the empirical distribution $P(x)$
immediately, but we cannot construct the coarsegrained distributions
$\{ P_i \}$ without first extracting the optimal partition $\{ l_i
\}$. In dynamical systems theory the optimal partition is provided by
the generating partition. The generating partition, if it exists, is
the signature of the dynamical system because it shows how the
dynamics coarsegrains phase space naturally. In contrast, the
histograms that appear on the optimal support can be produced by every
system in the same topologic universality class (symbolic dynamics is
universal for all systems in the same universality class
(\cite{gunaratne:90}, \cite{mccauley:chaos})), so that a particular
statistical distribution $\{ P_i \}$ cannot be the signature of a
particular dynamical system. Both the Henon map and the logistic map
$f(x) = D x (1-x)$, with $D_c < D < 4$, where $D_c$ is the period
doubling critical point, belong to the same topologic universality
class (both the logistic map with $D >4$ and the binary tent map
belong to a separate universality class). Both systems, although of
different spatial dimension, generate the same range of histograms
(for corresponding classes of initial conditions), but on different
supports. From the perspective of both dynamical systems theory and
the search for scale invariance the central problem of data analysis
is to extract the optimal partition of a particular set of data
points. See ref. \cite{glazier:88} and \cite{meneveau:87} for examples
of the extraction of optimal partitions from data in fluid
mechanics.

%%%%%%
\subsection{Multifractal scaling  (via the empirical distribution $P(x)$ 
on an efficient support)}
\label{sect:multi_empirical}

A nonuniform distribution on a fractal support looks
fractally-fragmented (looks more and more spiky as the support is
viewed with finer and finer resolution $l^{(n)}$). Distributions on
space-filling supports may also be fractally fragmented, as we shall
see below. A nonuniform distribution on a uniform or nonuniform
support (that is either fractal or space--filling) can be used to sort
and label fractal subsets of the support. Each subset has its own
fractal dimension $d(\alpha)$, where $\alpha$ is the labeling--index
\cite{halsey:86}. Multifractal, in this paper, always means a
spectrum of fractal dimensions, where each dimension describes the
scaling via (\ref{eq:1}) or (\ref{eq:1b}) of a subset of the support
of $P(x)$. We will show in parts \ref{sect:generating_functions} and
\ref{sect:lognormal} why the attempt to use other definitions leads to
confusion and failed expectations (predictions of ``dimensions'' that
are not the dimension of anything are discussed in parts
\ref{sect:correlation_integral} and \ref{sect:generating_functions}).

First, note that for a uniform distribution on a uniform support we
can write $P_i = N_n^{-1} = \left( l^{(n)} \right)^{D_0}$. To describe
a nonuniform distribution on a uniform or nonuniform support in
generation $n$ of coarsegraining, we try to define scaling exponents
$\alpha_i$ by writing $P_i = l_i^{a_i}$, where the scaling index
$\alpha_i$ takes on the same value $\alpha_i = \alpha$ on $N(\alpha)$
different intervals which we denote (using very sloppy but obvious
notation) by $l_1,\dots,l_{N(\alpha)}$. Therefore, by (\ref{eq:1b}),
we can write
\begin{equation}
\label{eq:1c}
\sum_{i=1}^{N(\alpha)} l_i^{d(\alpha)} \approx 1
\end{equation}
to define $d(\alpha)$ as the Hausdorff dimension of the subset of the
support where $\alpha_i = \alpha$ (if, indeed, such scaling holds). If
we could accurately replace the optimal nonuniform partition by the
largest term $l(\alpha) = \max \{ l_i : i = 1,\dots,N(\alpha) \}$ in
the subset, then we would obtain $N(\alpha) = l(\alpha)^{-f(\alpha)}$,
where $D_H > f(\alpha) > d(\alpha)$. In other words, $f(\alpha)$ is
the box--counting dimension for $N(\alpha)$ nonoverlapping uniform
intervals of size $l(\alpha)$. This replacement works only for
relatively uniform sub--clustering. Otherwise it is necessary to
compute $d(\alpha)$.

Given a nonuniform empirical distribution $P(x)$ over an optimal
uniform partition, whether or not the histograms $\{ P_i \}$ scale
over a reasonable range of different sizes of $l_i$, $P_i =
l^{\alpha_i}$, is the main question for empirical data analysis. In
cosmology $P_i = n_i/N$ is the fraction (out of a total number $N$ of
galaxies) of galaxies in the $i$th interval $l_i$.

In all data analysis and computer simulations there can be at most
finitely many values of $\alpha$ and $f(\alpha)$ (and finitely--many
values of $\lambda$ and $D(\lambda)$ as well) because $N_n \ll N$ is
finite ($N \approx 400$ for typical galaxy samples). However, the
number of points in a spectrum will grow generation by generation n
for a multifractal spectrum (again, this distinguishes multifractal
from bi--fractal, tri--fractal, $\dots$, $N$--fractal) within the cutoff
limits $l_{\rm max} \ge l^{(n)} \ge l_{\rm min}$. We can illustrate
this via a simple example of an $f(\alpha)$ spectrum, the one given by
the two--scale Cantor set with first generation probabilities $p_1 >
p_2$ describing statistical independence (\cite{halsey:86}) in all higher
generations, and optimal first generation intervals $l_1 = a^{-1}$ and
$l_2 = b^{-1}$. In this case, fixing the scaling index $\alpha$ picks
out a monofractal, so that $d(\alpha) = f(\alpha)$ because all
intervals in the subset have the same size $l_m = l_1^m l_2^{n-m}$. By
using Stirling's approximation on $n!$ (requiring $n \gg 1$), we then
obtain, with 
\begin{equation}
\alpha = \frac{-x \ln p_1 - (1-x) \ln p_2}{ x \ln a +(1-x) \ln b}, 
\end{equation}
that 
\begin{equation}
f(\alpha) \approx \frac{-x \ln x- (1-x) \ln (1-x)}{x \ln a + (1-x) \ln b}, 
\end{equation}
where $x = m/n = 0, 1/n, 2/n, \dots, 1$
parameterizes both $\alpha$ and $f(\alpha)$. There are $n+1$ points in
the spectrum so that $f_{\rm max}(a) < D_H$, where $a^{-D_H} +
b^{-D_H} = 1$ defines $D_H$. Note also that $f(\alpha) =
s(x)/\lambda(x)$, as expected.

We can summarize our present terminology by writing down either 
the generating function (\cite{halsey:86})
\begin{equation}
\label{eq:5}
\chi_n(q) = \sum_{i=1}^{N_n} P_i^q
\end{equation}
or the generation function (\cite{jensen:87})
\begin{equation}
\label{eq:6}
\Gamma_n(q) = \sum_{i=1}^{N_n} \frac{P_i^q}{l_i^\tau} \approx 1
\end{equation}
where, in the empirical search for scaling laws, it is first necessary
to find an approximately {\em optimal partition} in order correctly to
extract a multifractal spectrum of dimensions $f(\alpha)$, or even one
fractal dimension $D_H$. Otherwise the convergence requirements (the
number $n$ of generations in a hierarchy $\{ l_i \}$ of interval
sizes) needed to see scaling with an approximately correct exponent
almost always outruns the limitation placed by the number $N$ of
points in an empirical sample. 

When multifractal scaling can be shown to hold over enough different
generations $n$ of interval sizes $l^{(n)}$, then we can also write
\begin{equation}
\label{eq:5b}
\chi_n(q) \approx \sum_{\alpha} l(\alpha)_i^{q \alpha -f(\alpha)}
\end{equation}
and
\begin{equation}
\label{eq:6b}
\Gamma_n(q) \approx \sum_{\alpha} l(\alpha)^{q \alpha -f(\alpha) - \tau} 
\approx 1 .
\end{equation}
Note that (\ref{eq:6}) generalizes (\ref{eq:1b}) so that we can
explicitly discuss nonuniform distributions of points on the support,
as with (\ref{eq:5}).

By using $P_i = N_n^{-1}$ in (\ref{eq:6}) we get a result that looks
formally like the generating function $Z_n(\beta)$ in (\ref{eq:2})
above if we set $\beta = - \tau$ and $N_n^q = Z_n(\beta)$. Note,
however, that no assumption about the distribution of points $\{ P_i
\}$ over the support $\{ l_i \}$ was necessary in order to define the
generating function (\ref{eq:2}). In dynamical systems theory the
generating function $Z_n(\beta)$ can be rewritten via symbolic
dynamics as the partition function for a one dimensional Ising model
(\cite{jensen:87}) with long range interactions in equilibrium
statistical mechanics ($\beta$ is then the inverse temperature).

Most discussions of multifractals inevitably stress that the
generating functions (\ref{eq:5}) and (\ref{eq:6}) should {\em
themselves} scale approximately in the limit $l^{(n)} \rightarrow 0$
where, due to domination of the entire sum by $N(\alpha)$ largest terms
(\cite{halsey:86}, \cite{mandelbrot:91}), all of the same size (and
parameterized by $q$),
\begin{equation}
\label{eq:7}
\chi_n(q) \approx l(\alpha)^{q \alpha(q) - f(\alpha(q))}
\end{equation}
or
\begin{equation}
\label{eq:8}
\Gamma_n(q) \approx l(\alpha)^{q \alpha(q) - f(\alpha(q)) - \tau(q)}
\approx 1
\end{equation}
where (because $n$ goes to infinity as $l^{(n)} \rightarrow 0$) we
would hypothetically obtain an $f(\alpha)$ curve parameterized by
$q$. In this case, because there are enough points in the spectrum
that $f(\alpha)$ may be differentiated accurately, the ``generalized
dimensions'' $D_q$ can be defined by $\tau(q) = (q-1)D_q = q \alpha(q)
- f(\alpha(q))$, where $\alpha = \tau'(q)$ is the slope in the plot of
$\tau$ vs. $q$ and $q = f'(\alpha(q))$ is the slope of the $f(\alpha)$
curve. {\em This continuum limit is misleading because it is generally
inapplicable to data analysis.}

To see this, merely note that both generating functions are sums 
over all possible scales,
\begin{equation}
\label{eq:5c}
\chi_n(q) \approx l(\alpha_1)^{q \alpha_1(q) - f(\alpha_1(q))} + \dots +
l(\alpha_k)^{q \alpha_k(q) - f(\alpha_k(q))} 
\end{equation}
(or, in the case of a nonuniform distribution on a monofractal, over
$N_n$ terms $l^{\alpha_i}$ with different exponents $\alpha_i$). For
finite $l^{(n)}$ the generating functions (\ref{eq:5b}) and
(\ref{eq:6b}) {\em cannot} scale approximately unless $l^{(n)}$ is
small enough, $l^{(n)} \ll \ll 1$ (we cannot emphasize this
requirement too strongly), that a largest term approximation is
accurate, which is generally {\em not} the case. Formulations and
expectations of scaling based on the $l^{(n)} \rightarrow 0$ limits
(\ref{eq:7}) and (\ref{eq:8}) have been taken seriously enough to have
been employed during data analysis within the cosmology community
(\cite{jones:92}), as elsewhere. In data analysis this approximation
is usually a very bad one (see Theilor \cite{theilor:90} for a clear
and comprehensive exposition of the usual assumptions made in
discussions of multifractal generating functions).

In typical data analyses found in the literature the largest term
approximation is implicit in any plot of the logarithm of a generating
function vs. $\ln l^{(n)}$ in the search for generalized dimensions
$D_q$. We expect that most empirical data will not produce small
enough values of $l^{(n)}$ for a largest term approximation to be
applicable. {\em Even if multifractal scaling should hold term by term
in (\ref{eq:5}), in the form of (\ref{eq:5c}), it cannot be discovered
by a plot of $\ln \chi_n$ vs. $\ln l^{(n)}$.}  Instead, one must check
for scaling term by term inside the sum (\ref{eq:5}). {\em In other
words, forget the sums (\ref{eq:5}) and (\ref{eq:6}) and check each
term separately for scaling}, to verify whether $P_i = l_i^{\alpha_i}$
with $N(\alpha) = l_i^{-f(\alpha)}$ actually holds over a Geilo-range
of scale sizes. The generating functions are not directly measurable
anyway, so one needn't care whether or not they scale.

The coarsegrained density defined by $\rho_i = P_i/l_i =
l_i^{\alpha_i-1}$ is typically singular. Even if the support is
space--filling (even if $D_H = 1$) the coarsegrained density will look
more and more intermittent as the resolution is improved if the
distribution $P(x)$ has nonuniform clusters that are scale invariant
(see \cite{meneveau:87} for a one dimensional example from fluid
turbulence). Any attempt to replace a staircase $P(x)$ by a
distribution with a differentiable density may delete, mask, or, at
best, unnecessarily complicate the description of clustering and
intermittence. Why introduce the mathematical fiction of a continuous
distribution when observation gives us tractable discreteness
directly?

%%%%%%%%%
\section{The correlation integral}
\label{sect:correlation_integral}

\begin{figure}
 \begin{center} \epsfxsize=8cm 
 \begin{minipage}{\epsfxsize}\epsffile{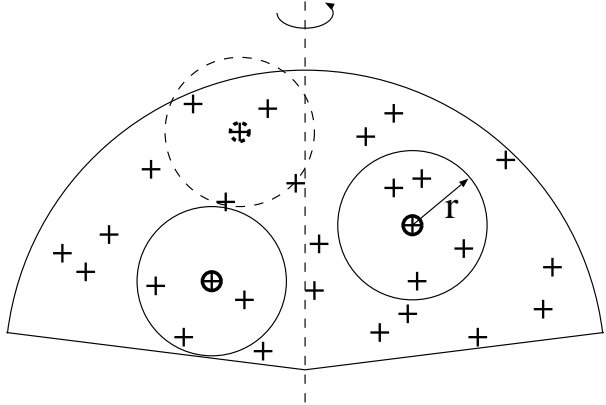}\end{minipage}
 \end{center}
\caption{ \label{fig:scal_method}
A sketch of the sample, illustrating which galaxies we use (solid
circle) and don't use (dashed circle) in the search for scale
invariance.}
\end{figure}

In cosmology (\cite{peebles:large}, \cite{coleman:fractal},
\cite{baryshev:94}), as it was in empirical analyses of dynamical
systems ten years ago (before the partitioning (\cite{feigenbaum:88})
and recycling (\cite{cvitanovic:88}) of strange sets), it is usual to
work with the correlation integral
\begin{equation}
\label{eq:9}
n(r) = \frac{1}{N} \sum_{i=1}^{N} n_i(r)
\end{equation}
where ``the correlation integrand''
\begin{equation}
\label{eq:10}
n_i(r) = \frac{1}{N} \sum_{i \ne j =1}^{N} \theta(r - | x_i - x_j|)
\end{equation}
is the fraction of galaxies in a sphere of radius $r$, centered on the
$i$th of $N$ galaxies. Here, we take $0 < r < 1$. This means that the
original dimensional variable $r$ for each galaxy has been rescaled by
dividing it by $r_{{\rm max},i}$, where $r_{{\rm max},i}$ is the value
of the unscaled variable $r$ for which the of radius $r$, centered on
galaxy $i$ just touches the boundary of the data set. In other words,
spheres of every radius $r$ lie {\em completely} within the boundaries
of the data (see figure~\ref{fig:scal_method}). We stress that data
sets should not be ``extended'' by adding points beyond the boundaries
of the observational data during box--counting. To do so would change
the data set from the one that we set out to analyze.  In other words,
we agree with the Pietronero school \cite{coleman:fractal}, but in
part \ref{sect:data} we will show how to refine the data analysis to
eliminate a certain (self--) inconsistency in that work.

Whenever the distribution in (\ref{eq:10}) exhibits statistical
independence then $n(r) = \chi_n(2)$ holds as well, as is implicit in
standard treatments. Clearly, data that are statistically independent
on an {\em optimal} partition (like the Cantor function differences
$P_i=2^{-n}$ over the closed intervals $l^{(n)}= 3^{-n}$) will not
show statistical independence over an {\em arbitrary} partition. In
the limit of small length scales the generalized dimension $D_2$
coincides with the correlation integral dimension $\nu$ for the case
of an empirical distribution $P(x)$ that exhibits statistical
independence on its optimal coarsegrained support
(\cite{hentschel:86}), but we cannot merely assume statistical
independence of observational data, and the zero length limit is
anyway physically unattainable. Therefore we do not expect to extract
$D_2$ via data analysis. Loosely speaking, however, one can refer to
$\nu$ as the ``correlation dimension''. Note that $\nu < 3$ requires a
either a multifractal distribution on a fractal support, or, for a
nearly uniform distribution, the support must be fractal.

The generalization of (\ref{eq:9}) is given by the (not directly
measurable) generating function
\begin{equation}
\label{eq:11}
G_n(q) = \frac{1}{N} \sum_{i=1}^N n_i(r)^{q-1}
\end{equation}
As in (\ref{eq:9}), $N$ is not the number of intervals in a
nonoverlapping efficient partition, but is the number of points in the
data sample. {\em The correlation integral was first emphasized in the
literature because it appears to allow us to circumvent the need to
find an optimal partitioning.} We will return to this point shortly.

Our first main point is that whether one uses (\ref{eq:5}),
(\ref{eq:6}) or (\ref{eq:11}) to study galaxy counts is irrelevant: a
generating function cannot scale unless all terms in the sum scale
separately, and only then if one term dominates. If scaling holds
locally but the data set is not monofractal, then each term in the
correlation integral (\ref{eq:9}) must have the form
\begin{equation}
\label{eq:12}
n_i(r) \approx r^{\nu(i)}
\end{equation}
where $\nu(i)$ is the local correlation integral index, the scaling
exponent (formally somewhat analogous to $\alpha_i$ in equation
(\ref{eq:5b}), except that here there is no accurate way to define a
spectrum of fractal dimensions $f(\nu)$ because an efficient partition
is not defined by (\ref{eq:11})) for galaxy counts for
$n=1,2,\dots,n_{\rm max}$ spheres with corresponding radii $r$ centered
on galaxy $i$. Clearly, in the absence of largest term dominance the
correlation integral representing local scaling with $N$ {\em different}
indices $\nu(i)$,
\begin{equation}
\label{eq:9b}
n(r) = \frac{1}{N} \sum_{i \ne j =1}^{N} c_i r^{\nu(i)} ,
\end{equation}
is not scale invariant because each term in the sum scales
differently: \mbox{$(\lambda r)^{\nu(i)} = \lambda^{\nu(i)}
r^{\nu(i)}$}. If, in a plot of $\log n(r)$ vs. $\log r$, linearity is
reported for a large enough range of values of $r$ that the result is
not spurious (see Figure \ref{fig:logplot}, then the likelihood is
that all terms inside the sum have approximately the same scaling
exponent $\nu(i) \approx \nu$, indicating that the data set is
approximately monofractal. For example, with $N$ points distributed
uniformly over the $N_n = 2^n$ intervals of the optimal partition
$l^{(n)} = 3^{-n}$ of the middle thirds Cantor set, the correlation
integral is dominated by a single term
\begin{equation}
\label{eq:13}
n(r) = n_i(r) = 2^{-n} - 2^{-N}
\end{equation}
and scales when $N \gg n$, $n(r) \approx 2^{-n} = r^{D_0}$ where $D_0
= ln2/ln3$ because $r = 3^{-n}$. In other words, $n(r)$ is
approximately scale invariant for $N \gg n$ because every term
$n_i(r)$ in the sum (\ref{eq:9}) is the same and is also scale
invariant. Here, $D_0=D_2=\nu$ holds because we have (implicitly
chosen to use) a uniform distribution on a monofractal. In particular
the analysis of part \ref{sect:data} shows, that one cannot assume
that $n_i(r) = r^\nu + \delta n_i(r)$, where $\delta n_i(r)$ is
Poisson noise.

In the analysis of empirical data, on the other hand, if scaling of
(\ref{eq:11}) is reported but has only been observed over a non--Geilo
range of values of $r$ then the resulting spectrum of generalized
correlation dimensions $\nu_q$ in $t(q) = (q-1)\nu_q$ defined by
\begin{equation}
\label{eq:14}
G_n(q) = \frac{1}{N} \sum_{i=1}^{N} n_i(r)^{q-1} \approx r^{t(q)}
\end{equation}
may be spurious (and appears only in the unphysical limits where $N
\rightarrow \infty$ and $r \rightarrow 0$). The variation of $t(q)$
with $q$ obtained from a plot of $\ln G_n(q)$ vs. $\ln r$ over an
inadequate range of $r$--values probably indicates that the generating
function (\ref{eq:11}) does {\em not} scale. In
\cite{aurel:92} it is shown how one can even get a spurious
``generalized dimension'' spectrum from log--log plots of a {\em
Gaussian} distribution.

{\em We point out next that the hope that one could circumvent the
need to extract the optimal partition from the empirical data was an
illusion:} the generating function (\ref{eq:11}) cannot be used to
compute either a Hausdorff or box--counting dimension. Setting $q=0$
in (\ref{eq:14}), the standard approach would lead the expectation
that $\nu_0$ in
\begin{equation}
\label{eq:15}
G_n(0) = \frac{1}{N} \sum_{i=1}^{N} n_i(r)^{-1} \approx r^{-\nu_0}
\end{equation}
provides an estimate for the box counting dimension $D_0$. This is
impossible, because neither the box counting nor information dimension
is included in the $\nu_q$ spectrum (appeals to the limit of vanishing
$r$ (\cite{sato:87}) do not help in the empirical case).  The reason is
simple: the terms on the left hand side of (\ref{eq:15}) don't define
an efficient, nonoverlapping partition of $G_n(0)$ intervals, each of
size $r$. Hence, the ``convergence'' difficulties reported by
\cite{martinez:clustering} in the attempt to estimate the box counting
dimension by computing $\nu_0$.

If we would try instead to define an interval $r_i$ by formally
writing $r^{\nu_0} (n_i(r)^{-1}) = r_i^d$ in (\ref{eq:15}), then the
result
\begin{equation}
\label{eq:15b}
\frac{1}{N} \sum_{i=1}^{N} r_i^d \approx 1
\end{equation}
reminds us superficially of equation (\ref{eq:1b}) above, but $d$ does
not define a Hausdorff dimension: the $N$ intervals $r_i$ overlap very
badly with each other because the sum is over all $N$ galaxies instead
of over an efficient nonoverlapping partition. Equation (\ref{eq:15b})
was proposed by Martinez (\cite{martinez:measures}) as one that yields
$D_H$, as well as the $D_q$ spectrum for $q< 1$ via the generalization
(see also \cite{martinez:proceed})
\begin{equation}
\label{eq:15c}
W_n(t) = \frac{1}{N} \sum_{i=1}^{N} r_i^{-t} \approx p^{-q}
\end{equation}
but information about the spectrum $D_q$, aside from an estimate 
of $D_2$, is not included in these formulae.

Equation (\ref{eq:15b}) is supposed to be based on the equation
\begin{equation}
\label{eq:16}
\frac{1}{N} \sum_{i=1}^{N} r_i \approx \overline{r_N}
\end{equation}
with
\begin{equation}
\label{eq:16b}
\overline{r_N} \approx K N^{-1/D}
\end{equation}
discussed by \cite{badii:84}, where $r_i$ is the nearest neighbor
distance between two points in a sample consisting of $N$
points. However, (\ref{eq:16}) is not the same as (\ref{eq:15b})
because the partitioning in (\ref{eq:16}) is {\em nonoverlapping}:
each point is connected only to one other point. For any finite subset
of the middle--thirds Cantor set consisting only of end points, we
find that $K^{-1} = 2$ because the number of end points $N = 2^{n+1}$
is simply twice the number of intervals $N_n = 2^n$ required to cover
those points. In this case the required nearest neighbor intervals are
simply the usual nonoverlapping intervals $r_i = 3^{-n}$ of the
middle--thirds Cantor set. Notice also that one cannot stray far from
the optimal result $l^{(n)} = 3^{-n}$ by using nonterminating ternary
expansions $x = .\epsilon_1 \dots \epsilon_N \dots$ with $\epsilon_i =
0$ or 2 to generate $N$ points of the middle--thirds Cantor set rather
than by using only terminating ternary strings. While scaling holds
{\em exactly} when we use nearest neighbor distances in (\ref{eq:16})
and (\ref{eq:16b}) for the mathematical idealization of a Cantor set,
we should not expect an analogous ``convergence rate'' when we use
empirical data. Equations (\ref{eq:16}) and (\ref{eq:16b}) should only
be expected to yield an accurate scaling index $D \approx D_H$ when
the nearest neighbor distances are taken to be the end points of an
optimal partition.

The expectation (\cite{martinez:90-2}, \cite{martinez:measures}) that
different generating functions can be used to compute ``the $D_q$
spectrum'' for different ranges of $q$, even in the idealized limit
where $l^{(n)} \rightarrow 0$, is based on three unfullfilled
expectations. First, the box counting dimension does not belong to the
$\nu_q$ spectrum (neither does the information dimension). Second, the
$\nu_q$ spectrum (defined by (\ref{eq:14}) in the limit of infinitely
small $r$) does not coincide with the $D_q$ spectrum of (\ref{eq:5b})
and (\ref{eq:6b}) (which do include the box counting and information
dimensions as $l^{(n)} \rightarrow 0$).  Third, one cannot change
probabilities and supports and expect scaling exponents to remain
invariant: neither multifractal spectra $f(\alpha)$, nor generalized
dimensions $D_q$ derivable from multifractal spectra, can be used to
define universality classes. The misconception that an optimal
partition is unnecessary has led to the expectation that partitions
can be manipulated without changing $f(\alpha)$ and
$D_q$. Multifractal spectra and generalized dimensions are
nonuniversal: they change whenever either the support or the
histograms on that support is changed. This is easily seen via the
simplest possible examples.

To emphasize this last assertion we demonstrate what happens when we
try to get a complete $D_q$ spectrum by combining results from
(\ref{eq:5}) and (\ref{eq:6}) for disjoint ranges of $q$, but while
using different distributions on different supports in each generating
function. The underlying point set is in each case taken to be a
finite number of points in the two--scale Cantor set. We consider
first a uniform partition with uneven probabilities ($p_1>p_2$ in
generation $n=1$) and statistical independence (for ease of
calculation), whereas in the second case we take even probabilities
$p_1 = p_2 = 1/2$ also with statistical independence on the uneven but
optimal partition with $l_1 > l_2$. We can cover the finite (or
infinite) point set in the first case by using an efficient uniform
(but not optimal) partition based on $l_1$, so that then we get
$D_\infty = \ln p_1/ \ln l_1$ and $D_{-\infty} = \ln p_2 / \ln
l_1$. If in the second case, we use (\ref{eq:6}) with $P_i = N_n^{-1}
= 2^{-n}$, then $D_\infty = - \ln 2 / \ln l_1$ and $D_{-\infty} = -
\ln 2 / \ln l_2$. The separate spectra do not lie on top of each other
because the end points $D_\infty$ and $D_{-\infty}$ do not
coincide. Clearly, we cannot combine these two different calculations
in order to estimate disjoint parts of the $D_q$ spectrum of either
case.

%%%%%%%
\section{Are galaxy distributions scale invariant?}
\label{sect:data}

In this search for scaling we use only the correlation ``integrand''
$n_i(r)$, not the correlation integral $n(r)$ for the reasons explained in
parts \ref{sect:multi_empirical} and \ref{sect:correlation_integral}.
We calculate the number of galaxies within a sphere of radius $r$
centered on each galaxy as depicted in Figure
\ref{fig:scal_method}. {\em We only use spheres that are completely
within the sample geometry}. Since we want to investigate the scaling
properties we are not allowed to apply any boundary corrections that
assume stationarity of the distribution of galaxies with respect to
translations (i.e.~homogeneity) or rotations (i.e.~isotropy). Such
``corrections'' would tend to introduce a spurious scaling with
dimension three. Boundary ``corrections'' are inherent in all the
usually used estimators for the two--point correlation function, see
e.g.~\cite{peebles:large}. Similar corrections were used for
estimators of the correlation integral see
e.g.~\cite{martinez:multiscaling}.
For the same reasons as above we use volume limited samples. Using
flux (or magnitude) limited samples one is usually using a weighting
scheme based on the selection function. In weighting with the
selection function one assumes homogeneity in giving a weight to
galaxies proportional to the mean density, which is determined mainly
from the nearby regions.

We only show galaxies in the plots where we have at least one
neighbour in a radius range larger than $\Delta_r$, our ``scaling''
range (with the limited data available it does not make sense to use
decades, since we only have at maximum one decade available). To
perform the scaling analysis more quantitatively we fit a straight
line to the $\log(n_i)$--$\log(r)$ plot and determine the slope, again only
for galaxies having a ``scaling'' range larger than $\Delta_r$.

We are severely limited by the small number density (or equivalently,
the small sample size) of the catalogues. Therefore we will not be
able to extend $\Delta_r$ over more than one decade which is obviously
too small to draw any conclusions about scaling (see
Figure~\ref{fig:logplot}). To estimate the influence of $\Delta_r$ on
the distribution of the slopes we look at three different
$\Delta_r$. To get a Geilo--range of scales from pie--shaped samples
(as in figure \ref{fig:scal_method}) we would need observational data
extending over several thousands \hMpc.

\begin{figure}
 \epsfxsize=14cm \epsfysize=20cm
 \epsffile{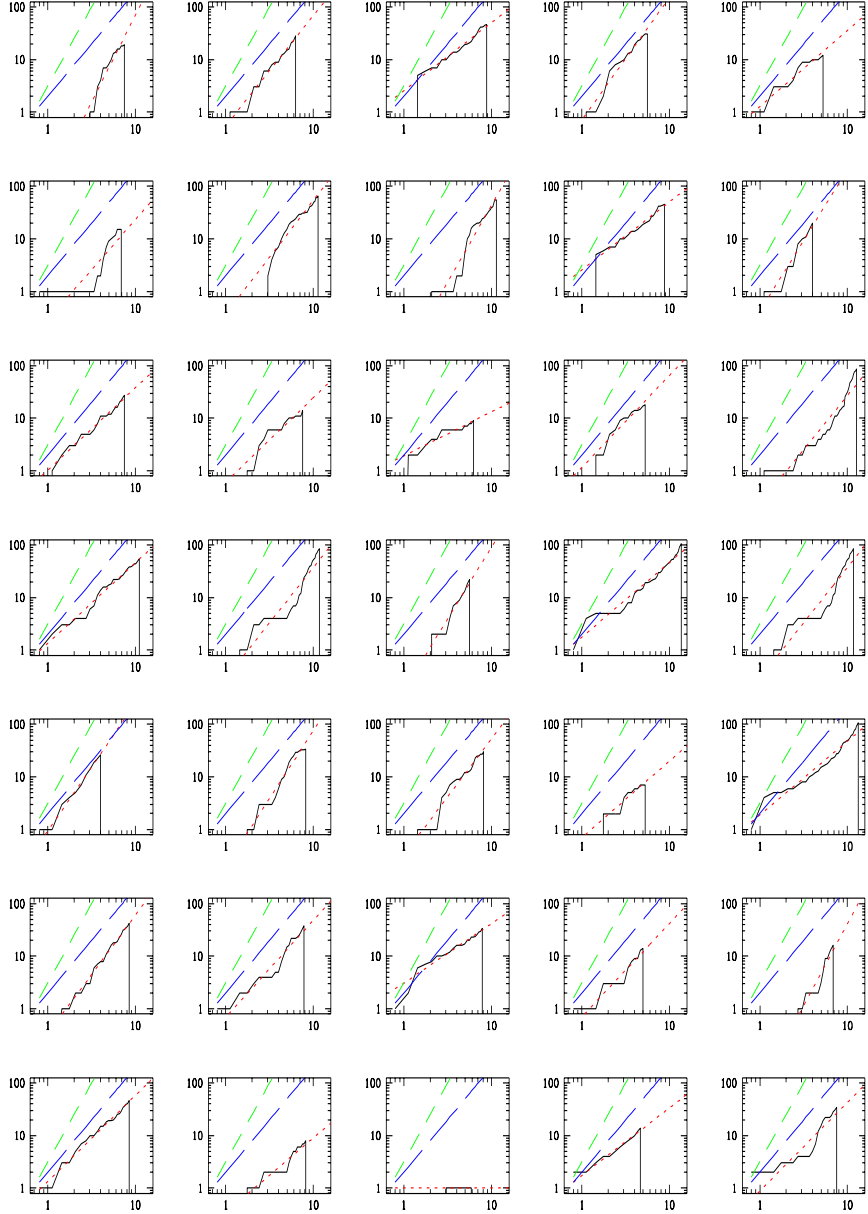}
\caption{ \label{fig:neigh_cfa_v4_N10}
Plots of $n_i(r)$ against $r$ (logarithmic axes) for volume limited
sample with 40~\hMpc\ depth (solid line) and "scaling" range $\Delta_r
\geq 3.1 \hMpc$. The dotted line is the fit, the long dashed line is
for $n_i(r) \propto r^2$ and the short dashed line for $n_i(r) \propto r^3$.}
\end{figure}
\begin{figure}
 \begin{center} \epsfxsize=14cm \epsfysize=20cm
 \begin{minipage}{\epsfxsize}\epsffile{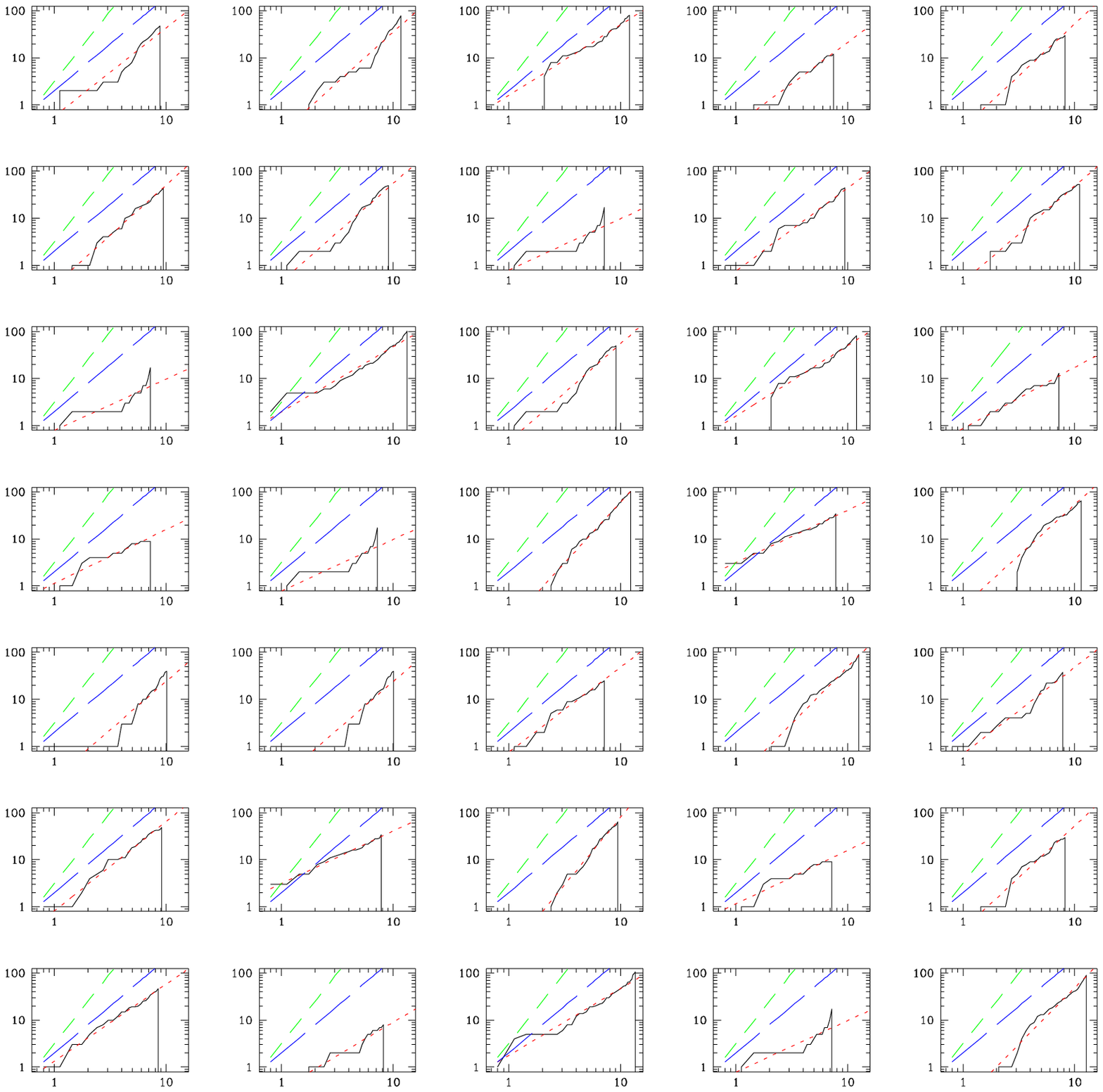}\end{minipage}
 \end{center}
\caption{ \label{fig:neigh_cfa_v4_N20}
Same as Figure~\ref{fig:neigh_cfa_v4_N10} but with "scaling" range $\Delta_r
\geq 6.2 \hMpc$.}
\end{figure}
\begin{figure}
 \begin{center} \epsfxsize=14cm \epsfysize=20cm
 \begin{minipage}{\epsfxsize}\epsffile{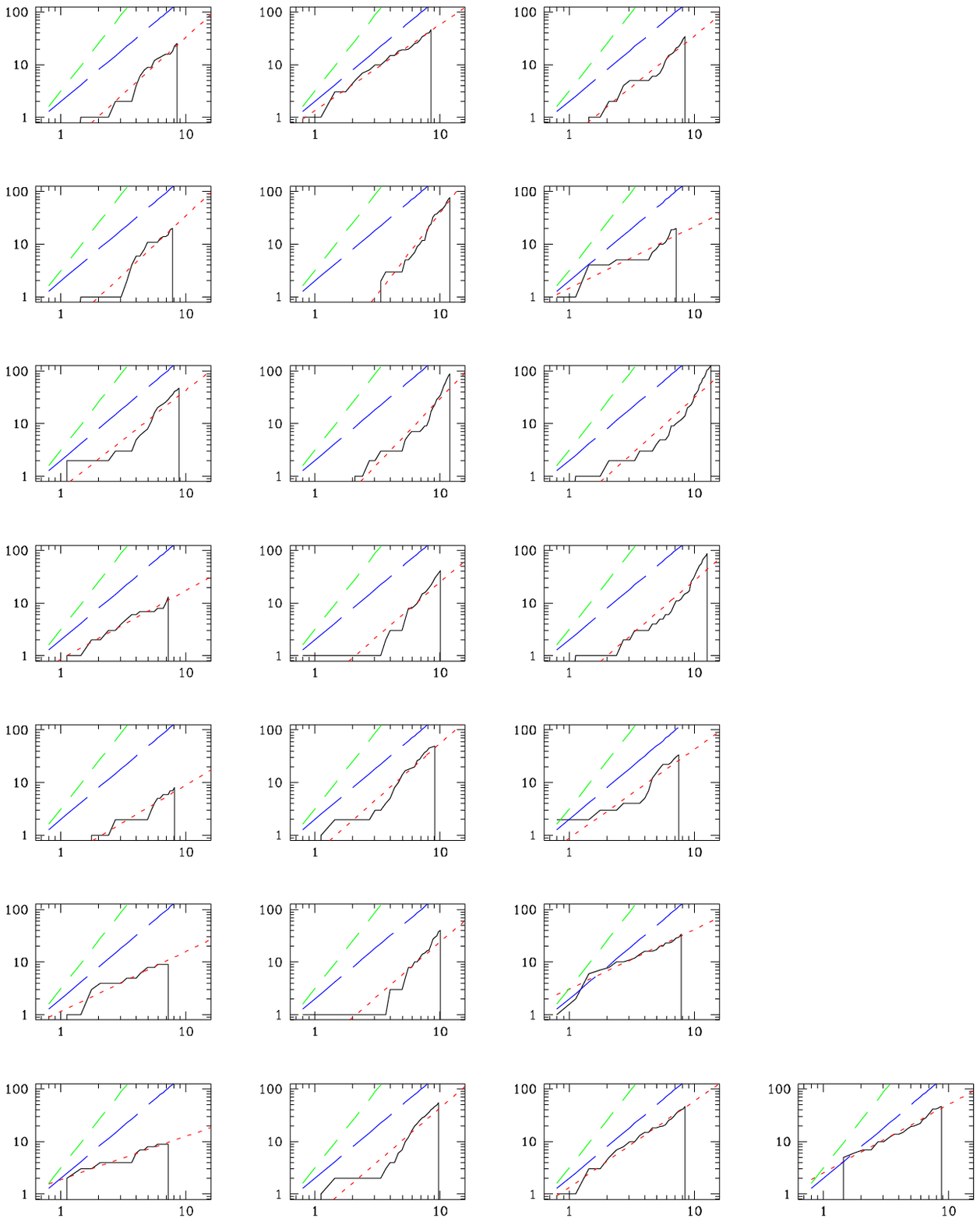}\end{minipage}
 \end{center}
\caption{ \label{fig:neigh_cfa_v4_N30}
Same as Figure~\ref{fig:neigh_cfa_v4_N10} but with "scaling" range $\Delta_r
\geq 9.3 \hMpc$.}
\end{figure}

\begin{figure}
 \begin{center} \epsfxsize=6.5cm
 \begin{minipage}{\epsfxsize}\epsffile{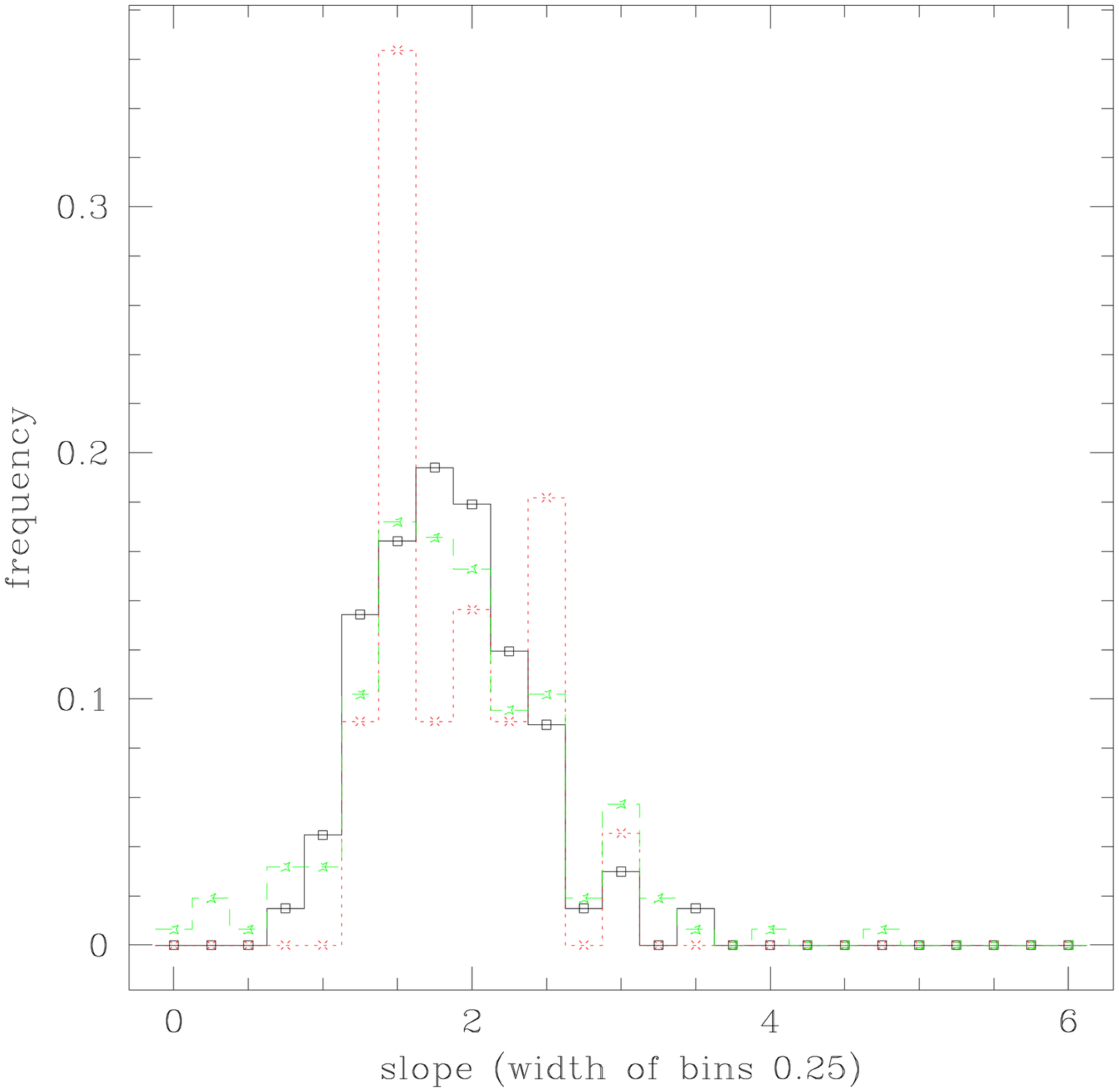}\end{minipage}
 \end{center}
\caption{ \label{fig:hist_cfa_v4}
The frequency of the slopes for the volume limited sample with
40~\hMpc\ depth, for the sample with $\Delta_r \geq 3.1 \hMpc$
(stars), with $\Delta_r \geq 6.2 \hMpc$ (open squares), and with
$\Delta_r \geq 9.3 \hMpc$ (crosses).}
\end{figure}
\begin{figure}
 \begin{center} \epsfxsize=6.5cm
 \begin{minipage}{\epsfxsize}\epsffile{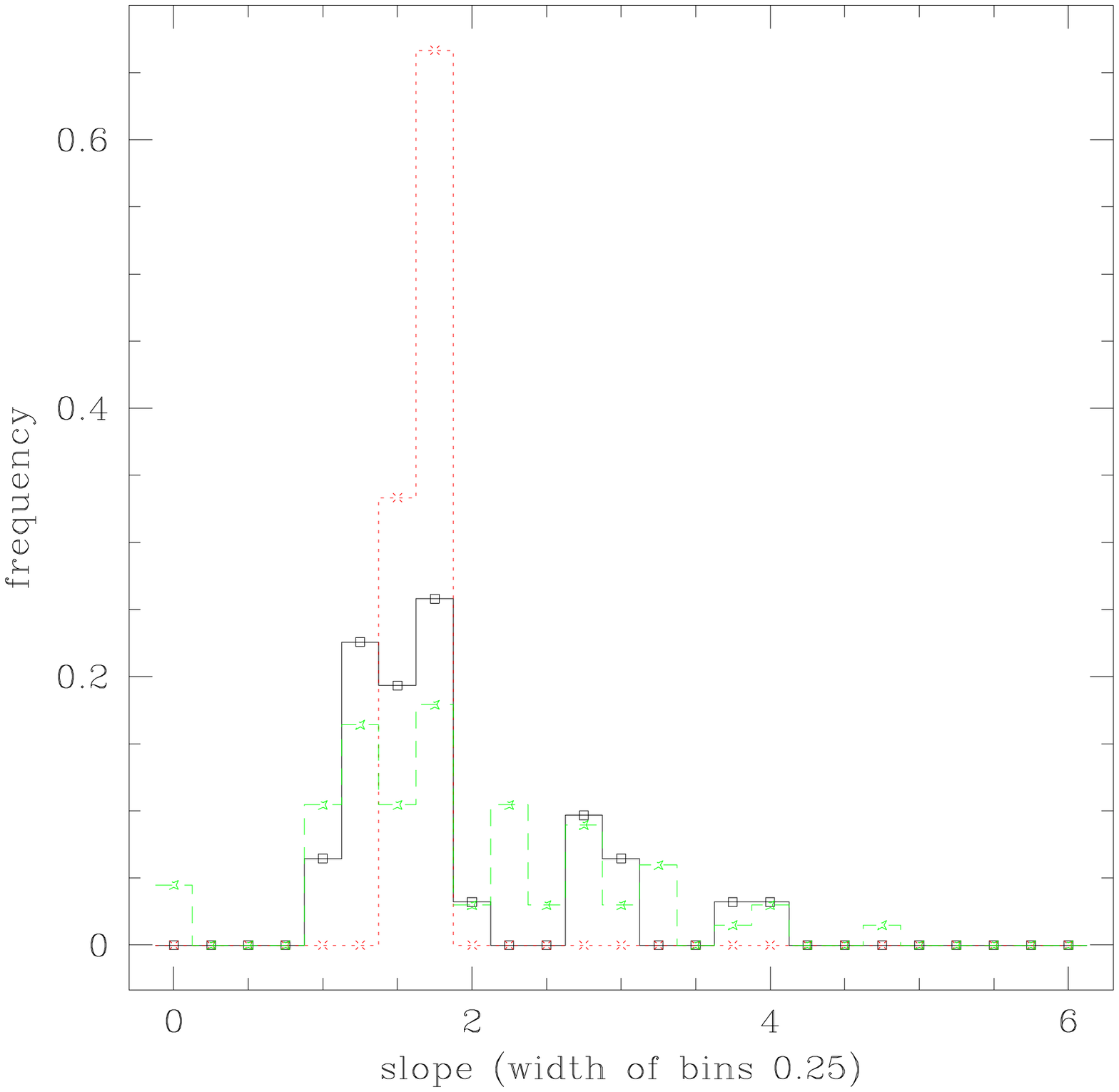}\end{minipage}
 \end{center}
\caption{ \label{fig:hist_cfa_v6}
The frequency of the slopes for the volume limited sample with
60~\hMpc\ depth, for the sample with $\Delta_r \geq 4.7 \hMpc$
(stars), with $\Delta_r \geq 9.4 \hMpc$ (open squares), and with
$\Delta_r \geq 14.1 \hMpc$ (crosses).}
\end{figure}
%

%%%%%%
\subsection{The CfA I galaxy catalogue} 
\label{sect:cfa_analysis}
We look, as an example of optically selected data, at the CfAI
catalogue a magnitude limited catalogue consisting of 1880 galaxies
(Huchra \cite{huchra:cfa1}). Using this data set fractal and
multifractal analysis were performed e.g. by Coleman \& Pietronero
\cite{coleman:fractal} or by Martinez \etal
\cite{martinez:clustering}.

First consider the volume limited sample with 40~\hMpc\ depth with
360 Galaxies in total, having a mean separation of 9.5 \hMpc. 
In Figure~\ref{fig:neigh_cfa_v4_N10} we show for 35 arbitrarily selected
galaxies the number of neighbours $n_i(r)$ against the radius of the
sphere $r$ for galaxies in the volume limited sample with 40~\hMpc
depth. In this case we demand a "scaling" range of $\Delta_r \geq 3.1
\hMpc$, resulting in 157 galaxies having a (more or less well) defined
slope.
In Figure~\ref{fig:neigh_cfa_v4_N20} we show the
scaling properties for 35 arbitrarily selected galaxies from the 67
galaxies with "scaling" range $\Delta_r \geq 6.2 \hMpc$, and in
Figure~\ref{fig:neigh_cfa_v4_N30} we show the scaling properties of
all the 22 galaxies which have a "scaling" range $\Delta_r \geq 9.3
\hMpc$, spanning roughly one decade.
As a first impression one recognizes that the scatter in the slope is
large and does not decrease for larger $\Delta_r$, which should
result in more reliable estimates for the slope.
To make this impression more quantitative we plot the frequency of the
slope for each of these samples in Figure \ref{fig:hist_cfa_v4}. One
has to bear in mind that this frequency table is in the case of $\Delta_r
\geq 9.3 \hMpc$ constructed from 22 galaxies only. In all three cases, 
the slope fluctuates over a range of 1.25 to 2.5. 

In Figure~\ref{fig:hist_cfa_v6} we plot the frequencies of the
slopes for the volume limited sample with 60~\hMpc, consisting of 215
galaxies with mean seperation 24~\hMpc. Imposing the constraints for
the "scaling" range we are left with 67 galaxies for $\Delta_r \geq
4.7 \hMpc$, 31 galaxies for $\Delta_r \geq 9.4 \hMpc$, and only 9
galaxies for $\Delta_r \geq 14.1 \hMpc$.
We see that no conclusions are possible with so few points. The peak
around 1.75 for $\Delta_r \geq 14.1 \hMpc$ (determined from only 9
galaxies!) is a mainly due to sampling galaxies from the same region
of the space in the center of the sample (see Figure~\ref{fig:scal_method}).
%

%%%%%%%%%%%%%%%%%%%%%%%%%%%%%%%%%%
\subsection{The IRAS 1.2 Jy galaxy catalogue}

Now we look at a sample of IRAS selected galaxies with limiting flux
1.2 Jy Fisher~\etal~\cite{fisher:irasdata} consisting of 5313
galaxies.  The big advantages of this sample is the nearly complete
covering of the sky, and the homogeneous flux calibration.
Fractal and multifractal analysis of IRAS galaxies were performed
e.g.~by Martinez \& Coles \cite{martinez:qdot}, Xia \etal
\cite{xia:fractal} and Labini \etal \cite{labini:numbercount}. The 
last mentioned authors claim, that the IRAS samples are too sparse to
estimate fractal dimensions reliably.

We proceed similar to the analysis of the CfA catalogue
\ref{sect:cfa_analysis}. As discussed in 
Kerscher~\etal~\cite{kerscher:fluctuations} we find differences
between the northern and southern hemisphere, but since we do not want
to focus on this topic we show the results for the combined data only.

\begin{figure}
 \begin{center} \epsfxsize=14cm \epsfysize=20cm
 \begin{minipage}{\epsfxsize}\epsffile{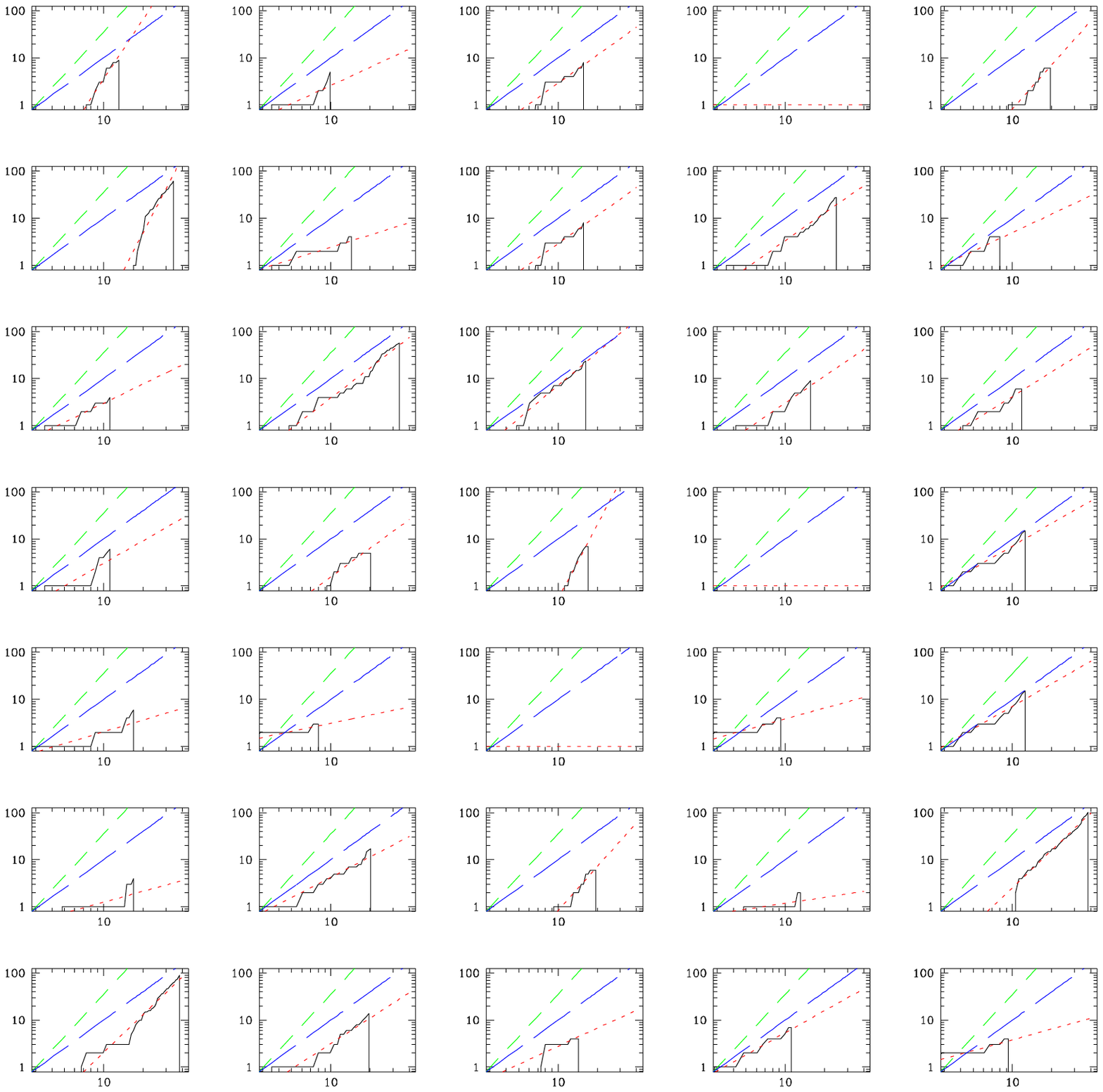}\end{minipage}
 \end{center}
\caption{ \label{fig:neigh_jy_v8_N10}
Plots of $n_i(r)$ against $r$ (logarithmic axes) for volume limited
sample with 80~\hMpc\ depth (solid line) and "scaling" range $\Delta_r
\geq 6.3 \hMpc$. The dotted line is the fit, the long dashed line is
for $n_i(r) \propto r^2$ and the short dashed line for $n_i(r) \propto
r^3$.}
\end{figure}

\begin{figure}
 \begin{center} \epsfxsize=14cm \epsfysize=20cm
 \begin{minipage}{\epsfxsize}\epsffile{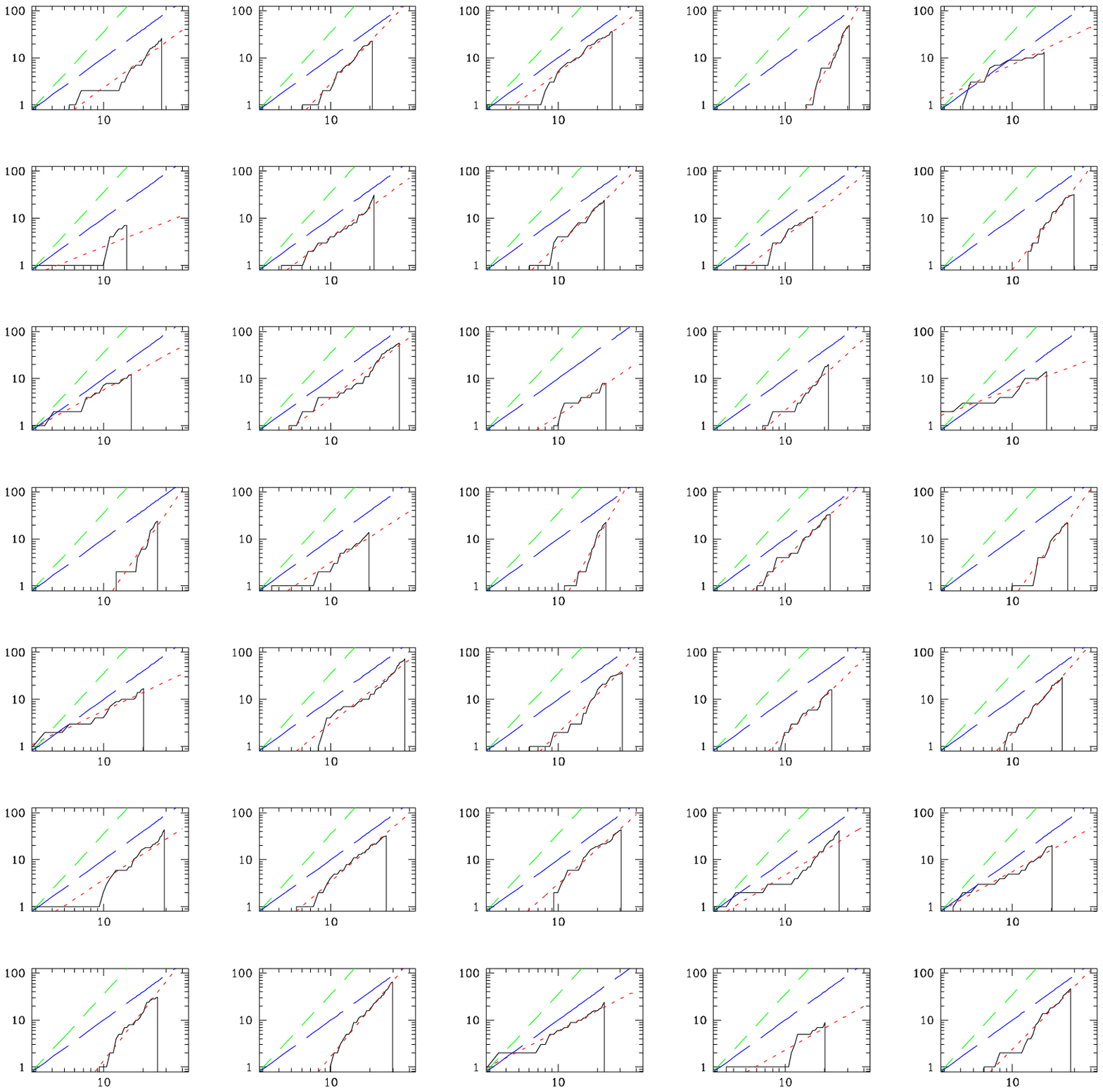}\end{minipage}
 \end{center}
\caption{ \label{fig:neigh_jy_v8_N20}
Same as Figure~\ref{fig:neigh_jy_v8_N10} but with "scaling" range $\Delta_r
\geq 12.6 \hMpc$.}
\end{figure}

\begin{figure}
 \begin{center} \epsfxsize=14cm \epsfysize=20cm
 \begin{minipage}{\epsfxsize}\epsffile{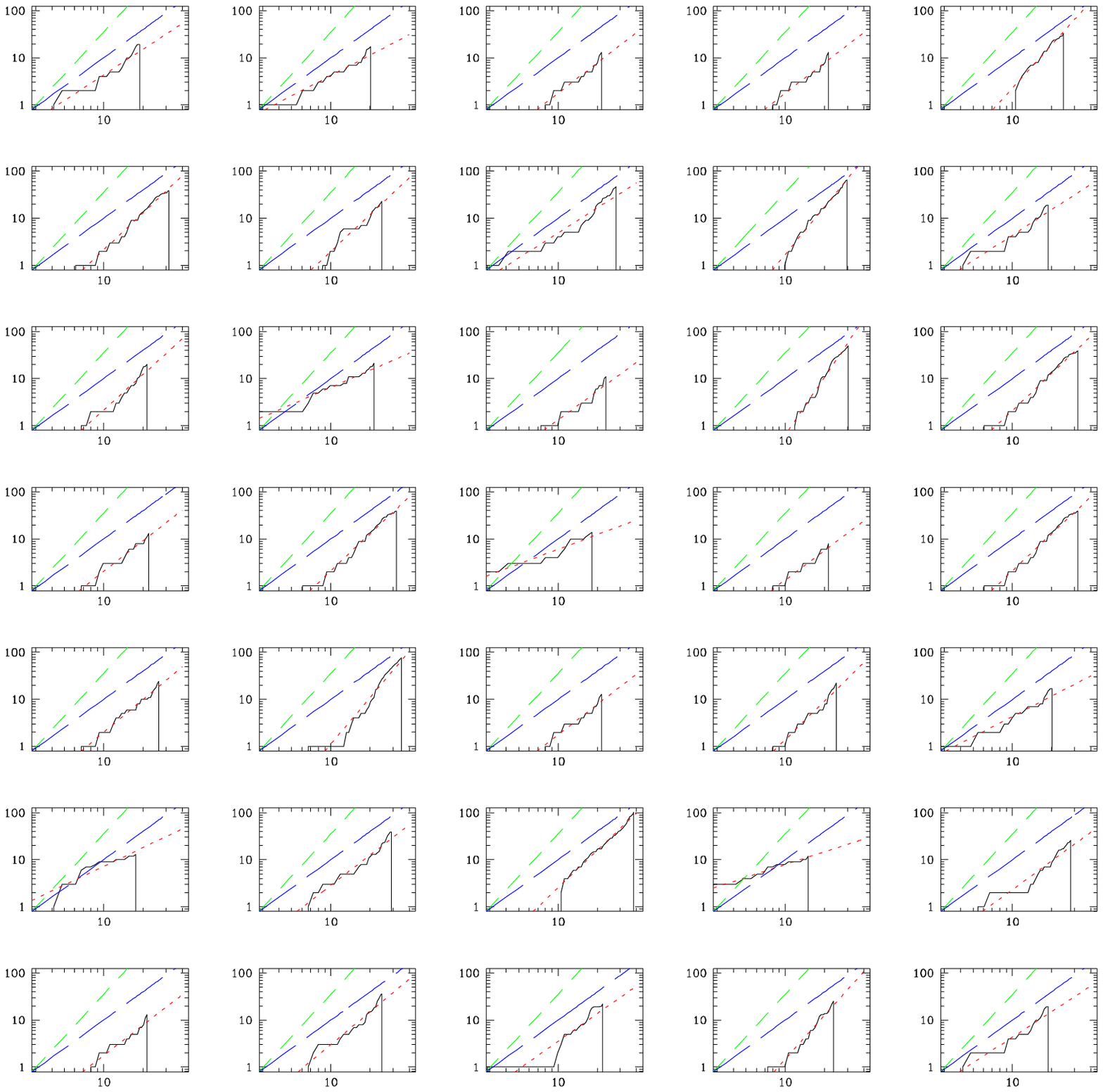}\end{minipage}
 \end{center}
\caption{ \label{fig:neigh_jy_v8_N30}
Same as Figure~\ref{fig:neigh_jy_v8_N10} but with "scaling" range $\Delta_r
\geq 18.9 \hMpc$.}
\end{figure}

In Figure~\ref{fig:neigh_jy_v8_N10} we show for 35 randomly selected
galaxies the number of neighbours $n_i(r)$ against the radius of the
sphere $r$ for galaxies in the volume limited sample with 80~\hMpc\
depth (mean seperation of the galaxies is 24 \hMpc). Restricting
ourselves to a "scaling" range of $\Delta_r \geq 6.3 \hMpc$ we are left
with 359 galaxies of the total 788 galaxies in the volume limited
sample.
In Figure~\ref{fig:neigh_jy_v8_N20} we show the scaling properties for
35 randomly selected galaxies from the 167 galaxies with a "scaling"
range of $\Delta_r \geq 12.6 \hMpc$, and in
Figure~\ref{fig:neigh_jy_v8_N30} 35 randomly selected galaxies from
the 72 galaxies with a "scaling" range of $\Delta_r \geq 18.9 \hMpc$.
Again we have a "scaling" range only spanning roughly one decade.

The sample with $\Delta_r \geq 6.3 \hMpc$ is clearly inapropriate for
an analysis, since often only one neighbouring galaxy is within the
scaling range, giving rise to a spurious slope of zero.
Again we see a large scatter in the slopes for all $\Delta_r$, which
does not decrease if we got to higher $\Delta_r$.


\begin{figure}
 \begin{center} \epsfxsize=6.5cm
 \begin{minipage}{\epsfxsize}\epsffile{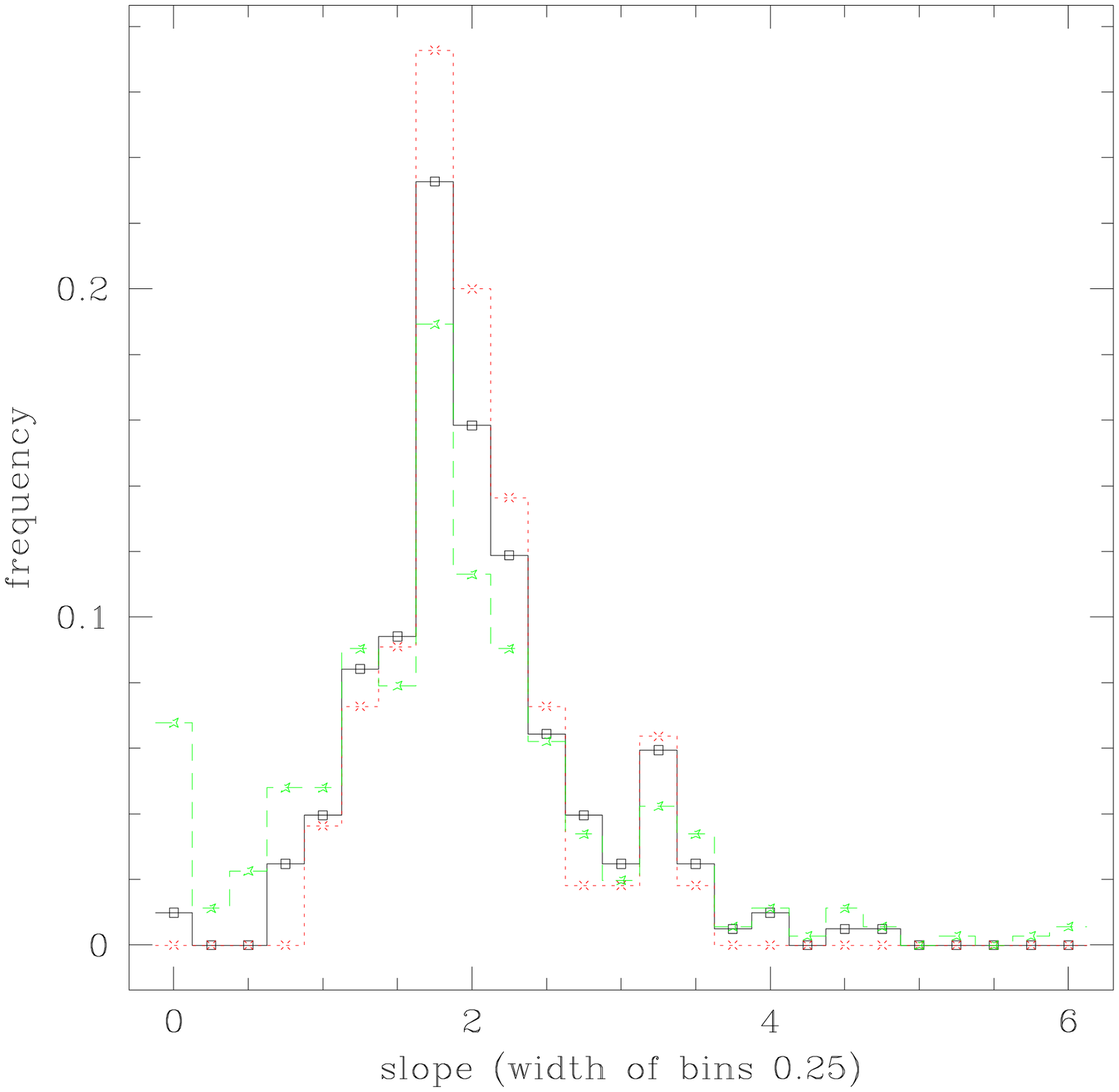}\end{minipage}
 \end{center}
\caption{ \label{fig:hist_jy_v4}
The frequency of the slopes for the volume limited IRAS sample with
40~\hMpc\ depth, for the sample with $\Delta_r \geq 3.1 \hMpc$
(stars), with $\Delta_r \geq 6.2 \hMpc$ (open squares), and with
$\Delta_r \geq 9.3 \hMpc$ (crosses).}
\end{figure}

\begin{figure}
 \begin{center} \epsfxsize=6.5cm
 \begin{minipage}{\epsfxsize}\epsffile{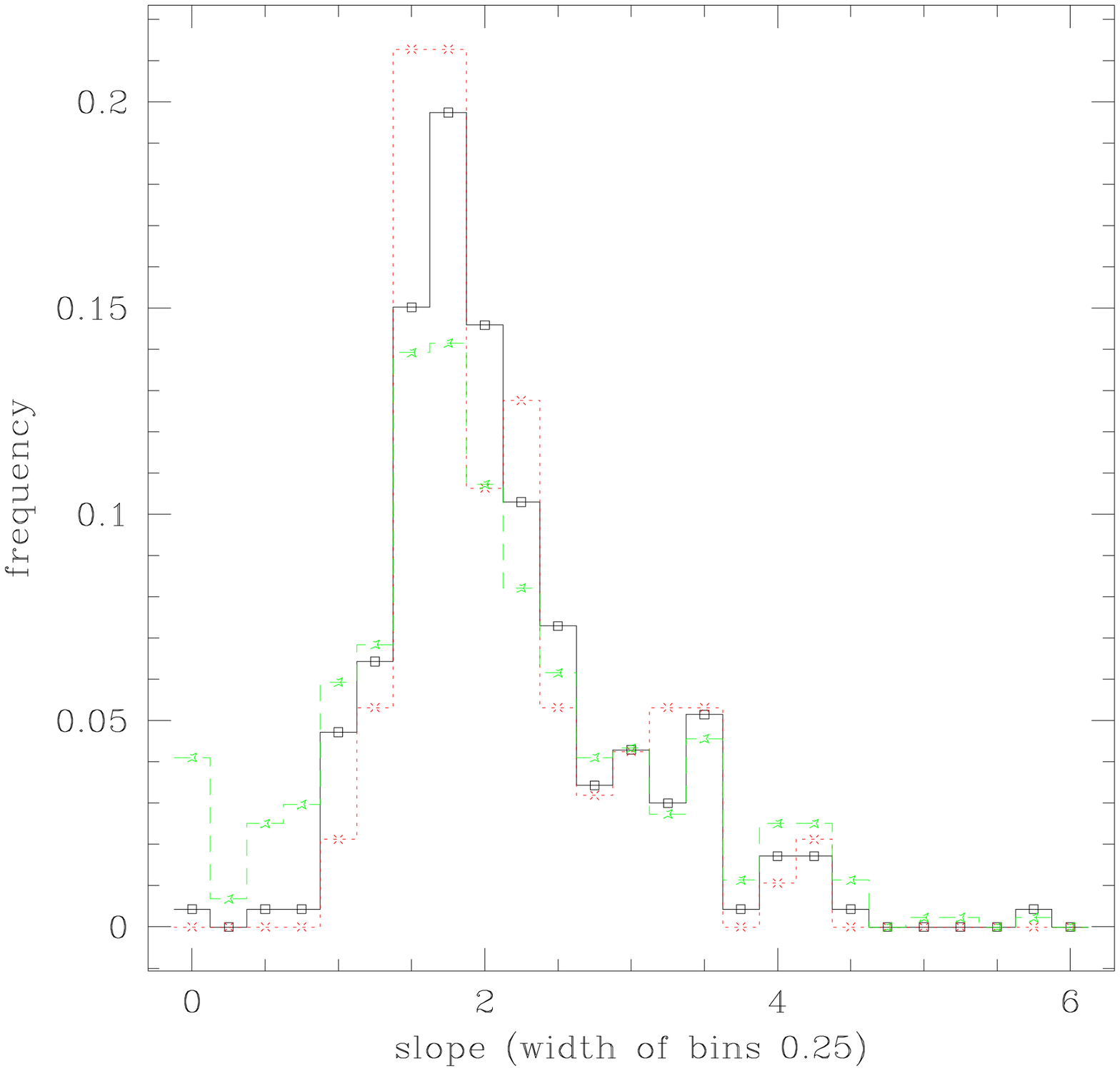}\end{minipage}
 \end{center}
\caption{ \label{fig:hist_jy_v6}
The frequency of the slopes for the volume limited IRAS sample with
60~\hMpc\ depth, for the sample with $\Delta_r \geq 4.7 \hMpc$
(stars), with $\Delta_r \geq 9.4 \hMpc$ (open squares), and with
$\Delta_r \geq 14.1 \hMpc$ (crosses).}
\end{figure}

\begin{figure}
 \begin{center} \epsfxsize=6.5cm
 \begin{minipage}{\epsfxsize}\epsffile{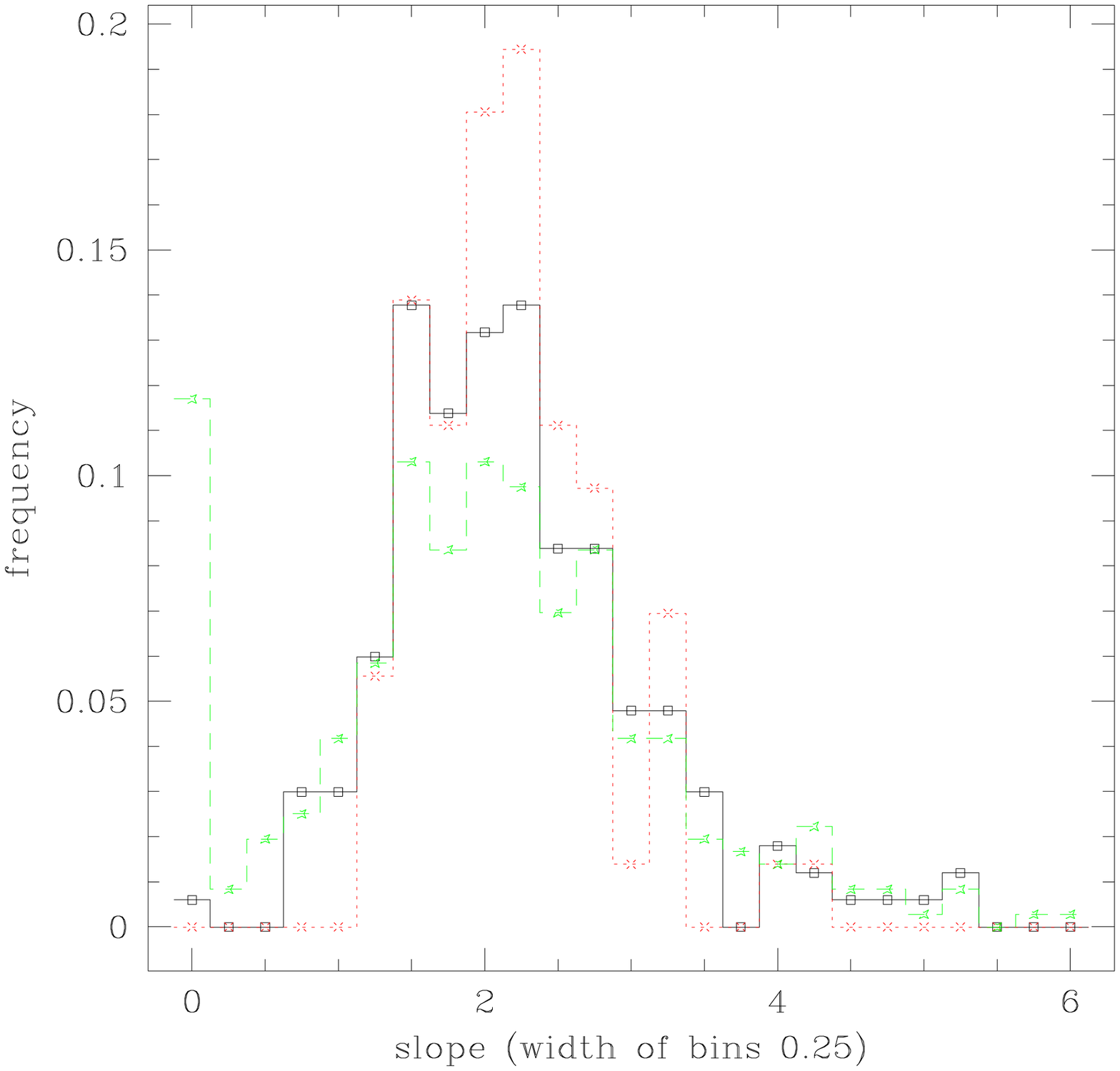}\end{minipage}
 \end{center}
\caption{ \label{fig:hist_jy_v8}
The frequency of the slopes for the volume limited IRAS sample with
80~\hMpc\ depth, for the sample with $\Delta_r \geq 6.3 \hMpc$
(stars), with $\Delta_r \geq 12.6 \hMpc$ (open squares), and with
$\Delta_r \geq 18.9 \hMpc$ (crosses).}
\end{figure}

To make this more quantitative we again plot histograms of the slopes,
now for a sequence of volume limited samples. The sample with
40~\hMpc\ depth containing 646 galaxies is shown in
Figure~\ref{fig:hist_jy_v4}. The sample with 60~\hMpc\ depth
containing 880 galaxies is shown in Figure~\ref{fig:hist_jy_v6}. The
sample with 80~\hMpc\ depth with 788 galaxies is shown in
Figure~\ref{fig:hist_jy_v8}.

%

%%%%%%%%%%%%
\subsection{Discussion of the results}
In both catalogues the local scaling exponents $\nu(i)$ fluctuate over
a broad range (see Figures \ref{fig:hist_jy_v4}--\ref{fig:hist_jy_v8}
and Figures \ref{fig:hist_cfa_v4},
\ref{fig:hist_cfa_v6}) indicating that there is {\em no} global
monofractal scaling. It mainly tells us that the distribution admits
large fluctuations. From the limited data we are not able to judge
whether these fluctuations are scale invariant (over three decades) or
not.
Therefore, fractal scaling, and certainly multifractal scaling, cannot
be deduced from that limited data as claimed in e.g.~
\cite{coleman:fractal}, \cite{martinez:clustering}, \cite{martinez:qdot}, 
\cite{xia:fractal}, \cite{labini:numbercount}. The fact that the 
correlation integral (or the two point correlation function)
apparently scales with one exponent is in this case not related to the
scaling of the galaxy distribution (again, see Figure \ref{fig:logplot}). We
want to emphasize that the broad range of different slopes is {\em
not} a sign of multifractality, it only shows that we have large
fluctuations.
In e.g.~\cite{pietronero:96} the authors claim to see scale invariance
over three decades (from 1 \hMpc\ to 1000 \hMpc). This statement is
based on the scaling properties of only one $n_i(r)$. In this case
$n_i(r)$ should be the number of galaxies in a sphere centered on our
galaxy.  However, the analysis is carried out by using only
intersections of a cone with a sphere as discussed in
\cite{labini:numbercount}. Such an analysis is inconsistent with their
own definition of fair sampling, very badly violating Fig.~8 of
\cite{coleman:fractal}.

In fitting straight lines to a log--log plot over roughly one decade
we have performed a superficial data analysis, one that gives
excessive weight to galaxies near the center of the
sample. Unfortunately nothing more is possible without postulating
``boundary corrections'' that have no basis in observation, or without
even more unfairly weighting points near the center of a conical
sample more heavily than those near a boundary.  But even in doing so,
we do not find any indication of global scaling. The main message is
that the current data are insufficient for a reliable scaling
analysis. We end this section with a quote from
\cite{coleman:fractal}:
``\dots if a sample contains too few points there may be no way to get
any information from it. In such a case one has to wait for better
(observational) data.''

%%%%
\section{Generating functions}
\label{sect:generating_functions}

The term multifractal is defined in Jones \cite{jones:origin} by
requiring only that the moments of an arbitrary distribution $P(X,l)$,
\begin{equation}
\label{eq:17}
\langle X^{q-1} \rangle = \sum_{X} P(X,l) X^{q-1} ,
\end{equation}
where $X$ is a random variable defined in or on intervals of size 
$l$, should scale,
\begin{equation}
\label{eq:17b}
\langle X^{q-1} \rangle \approx l^{\zeta_q},
\end{equation}
for small enough interval sizes $l$. However, without further requirements on
$X$ and $P(X,l)$, there is no reason to expect that scaling exponents
in (\ref{eq:17b}), if they exist {\em at all} for a given distribution
$P(X,l)$, bear any relation to generalized dimensions $D_q$ derivable
in the infinite precision limit from multifractal spectra $D(\lambda)$
or $f(\alpha)$. To be specific, is the partition nonoverlapping?
Efficient?  What is fractal about $X$ or $P(X,l)$? We will show via
example in part \ref{sect:lognormal} that equation (\ref{eq:17}) with
a scaling law (\ref{eq:17b}) generally does not describe intermittence
due to voids, and that if one tries in those cases to force the
definition $\zeta_p = (p-1)D_p$ (or, as in \cite{jones:origin},
$\zeta_p = (1-p) D_p$) then the $D_p$ are not dimensions
describing the support of the probability distribution, or of any
other coarsegrained set or subset connected with an arbitrary
distribution $P(X,l)$.

The generating function (\ref{eq:17}) combined with the scaling
expectation (\ref{eq:17b}) is used in definitions of {\em multiaffine}
fractals (\cite{barabasi:fractal}), where a deterministic or random
variable (or field) $X$ is continuous but has singular (or no)
derivatives. If the distribution of singularities of the field $X$ can
be described locally by writing $X \approx l^h$ and $P(X,l) \approx
l^{-f(h)}$, then (\ref{eq:17}) may or may not yield scaling exponents
$\zeta_p$ that give rise to a spectrum of generalized dimensions,
defined by $(p-1)D_p = p h(p) - f(h(p))$ in the limit $l
\rightarrow 0$, where $D_0, D_1$ and $D_2$ really are {\em fractal
dimensions} of {\em something} in the model. This happens {\em only}
when $f(h)$ describes a spectrum of fractal dimensions. Otherwise $h$
and $f(h)$ are just a rewriting of a nonfractal probability
distribution $P(X,l)$ via a coordinate transformation, and {\em
nonfractal distributions cannot be made fractal (or the converse) by a
differentiable coordinate transformation}. Stated another way, $f(h)$
is not a spectrum of Hausdorff dimensions unless ``$l$'' represents an
optimal or at least efficient partition of the support of an
underlying pointwise distribution $P(x)$. Examples in the literature
where $X \approx l^h$ with $P(X,l) \approx l^{-f(h)}$ are used without
any requirement of efficient partitioning of a support are height
fluctuations in surface roughening (\cite{barabasi:fractal}),
self--organized criticality (\cite{kadanoff:89}) and velocity
structure functions in the inertial range of fluid turbulence
(\cite{frish:turbulence}). In contrast with multiaffine fractals
(where there is no idea of a generating partition), for a
self--similar fractal (heretofore called ``fractal'') the point set is
generally spatially-fragmented (Koch and Peano curves are, however,
continuous), like a Cantor set, and $P_i$ scales like $l_i^{\alpha_i}$
({\em not} like $l_i^{-f(\alpha_i)}$) in order to describe a highly
singular density $\rho_i = P_i/l_i = l_i^{\alpha_i-1}$ on the optimal
partition describing the support of $P(x)$.

In a spirit similar to the attempt to define multifractal by using
the moments (\ref{eq:17}) of an arbitrary distribution $p(X,l)$ with a
scaling law (\ref{eq:17b}), the definition
\begin{equation}
\label{eq:18}
\langle P^{q-1} \rangle = \sum_{P} p(P,l) P^{q-1} ,
\end{equation}
where the probability distribution $p(P,l)$ is undefined, is treated
in various places (\cite{columbi:92}) as if it would be identical with
the generating function
\begin{equation}
\label{eq:5c_2}
\chi_n(q) = \langle P^{q-1} \rangle_{\rm coarsegrained} = 
\sum_{i=1}^{N_n} P_i  P_i^{q-1} ,
\end{equation}
although it is not.

For an arbitrary probability distribution $p(P,l)$ these two
generating functions are not even related; their scaling exponents (if
scaling exponents exist in either case) are not necessarily the same
even if the generating functions {\em are} qualitatively related. {\em
Multifractal spectra and generalized dimensions are not
universal}. Instead, they change with the histograms and their
support. Furthermore, any deviation from the empirical distribution
$P(x)$ and its optimal coarsegrained descriptions $\{ P_i \}$ is
equivalent to changing the underlying data set. Rather than imagining
that the empirical distribution $P(x)$ can be treated as a random
field that fluctuates from one galaxy sample to the other (as in Jones
\cite{jones:nonlinear}), we view different $P(x)$'s from different
samples as disjoint pieces of a {\em single} global distribution of
galaxies whose local properties can be discovered empirically, but
whose entire global (one hesitates to say ``universal'') aspect can
never be known due to the inherent limitations on observation. We do
not want to try to replace what we do not know ($P(x)$ {\em measured}
globally) with speculations that cannot be tested ($p(P,l)$ {\em
postulated} globally).
The notion of statistical ensembles is useless here: There is only one
universe, and it is not in equilibrium.

The source of confusing together entirely different generating
functions can be traced to the use of the infinite precision limit in
papers on dynamical systems theory written more than ten years ago. It
is suggested in \cite{hentschel:86} (see also \cite{grassberger:88})
that (\ref{eq:5c_2}) is analogous to the Lebesgue integral
\begin{equation}
\label{eq:19}
\langle P(B_l(x))^{q-1} \rangle = \int \d P(x) P(B_l(x))^{q-1}
\end{equation}
and should yield the same generalized dimensions $D_q$ in the infinite
precision limit, where $P(x)$ is supposed to be ``the natural
invariant measure'' of a chaotic dynamical system on a strange
attractor (see \cite{gunaratne:88}, e.g., for one definition of
``natural'') and $P(B_l(x)$) is the fraction of points lying within a
ball of size $l$ covering (but not necessarily centered on) a data
point $x$.  There are two serious difficulties with the attempt to
replace (\ref{eq:5c_2}) by (\ref{eq:19}), and both revolve about lack
of uniqueness.

First, there is no evidence from observation that a chaotic dynamical
system generates ``a natural measure'' for the various initial
conditions (meaning also ``present conditions'') found in nature.
Mathematically seen, a chaotic dynamical system can generate
infinitely many different distributions (``measures'') $P(x)$ for
infinitely many different classes of initial conditions
(\cite{mccauley:chaos} \& \cite{mccauley:classical}).  Empirically,
the dependence on initial conditions is not a problem: the data are
described by the empirical staircase
\begin{equation}
\label{eq:3b_2}
P(x) = \frac{1}{N} \sum_{i=1}^{N} \Theta(x-x_i) .
\end{equation}
Without having made any theoretical assumptions that prejudice the
data analysis we can say that the initial conditions, whatever they
were, produced the empirical distribution $P(x)$ via the time
evolution of some dynamical system.

Given the empirical measure (\ref{eq:3b_2}) there is still ambiguity
inherent in the attempt to use (\ref{eq:19}) as a replacement for
(\ref{eq:5c_2}). In finite precision there are different possible
definitions of the integral, depending on which subset of the data set
we decide to measure (before we can identify the function $P(B_l(x))$
we must first define ``$l$'').

If we choose the balls/intervals $B_l(x)$ to have arbitrary length
$l$, centered on a data point $x_i$ (as in \cite{pawlezik:87}), then
the fraction of points lying within each interval of size $l$ is given
by $P(B_l(x)) = n(x,l)$ where $n(x_i,l) = n_i(l)$ is the correlation
integrand
\begin{equation}
\label{eq:10_2}
n_i(r) = \frac{1}{N} \sum_{i, j =1; i \ne j}^{N} \theta(l - | x_i - x_j|) .
\end{equation}
Deleting the term $j=i$ in (\ref{eq:10_2}) is unimportant if the
intervals are large enough to give ``good statistics'' (pedantically,
one can also replace the factor $1/N$ by $1/(N-1)$ in
(\ref{eq:10_2})). Insertion of the pointwise definitions $P(B_l(x)) =
n(x,l)$ and
\begin{equation}
\label{eq:3_2}
\d P(x) = \frac{\d x}{N} \sum_{i=1}^{N} \delta(x-x_i)
\end{equation}
into the integral (\ref{eq:19}) yields the correlation integral
generating function
\begin{equation}
\label{eq:11c}
\int \d P(x) P(B_l(x))^{q-1} = \frac{1}{N} \sum_{i=1}^{N} n_i(l)^{q-1} 
= G_n(q) ,
\end{equation}
which differs significantly from (\ref{eq:5c_2}) in data analysis, as we
have emphasized in section \ref{sect:correlation_integral}.

In dynamical systems theory the $N$ intervals can in principle be
chosen small enough not overlap with each other: on a
mathematically--defined strange attractor there are $t^\infty$ points
in any neighborhood of any arbitrary point $x_i$ on the attractor
($t^\infty$ is the cardinality of the attractor). Here, the $N$
intervals (or balls) $B_l$ of size $l$ can be chosen small enough not
to overlap, but certainly do not partition the attractor efficiently,
if at all. Equation (\ref{eq:11c}), which was not invented with
partitioning in mind, is merely a time average over $N$ points on the
attractor, and the uniform weight $1/N$ is correct because each point
$x_i$ occurs exactly once (so long as trajectories of the dynamical
system are unique, which we assume here). In nonlinear dynamics
calculations the number $N$ of points may be increased by increasing
the precision of the calculation. In cosmology, in contrast, $N$ is
the total number of galaxies in a finite sample, so that the $N$
intervals of size $l$ are always overlapping.

There is a different way to define balls $B_l(x)$ and a corresponding
function $P(B_l(x))$. Instead of choosing $N$ uniform intervals where
$N$ is the number of data points, requiring the pointwise definition
$P(B_l(x)) = n(x,l)$ as given by (\ref{eq:10_2}) (which does {\em not}
include a partition of the data set), we can instead choose our balls
$B_l$ to be the $N_n$ intervals $\{ l_i \}$ in the optimal partition
of the data. Then, $P(B_l(x))$ is given by the simple function
\begin{equation}
\label{eq:20}
P(B_l(x)) = \sum_{i=1}^{N_n} P_i \chi_{l_i} (x)
\end{equation}
where $\chi_{l_i} (x)$ is the characteristic function for the
partition $\{ l_i \}$ of disjoint intervals \cite{royden:real} and
\begin{equation}
\label{eq:21}
P_i = P(x_i + l_i) - P(x_i) = 
\sum_{j=1}^{i+n_i -1} \theta(x_{i+n_i} - x_j) = n_i/N
\end{equation}
Here, $x_i$ and $x_{i+n_i} = x_i+l_i$ are taken to be the end points
of any of the $N_n$ {\em optimal} intervals $\{ l_i \}$. With this
optimal choice of ``what to measure'' (optimal choice of function
$P(B_l(x))$ to integrate with respect to the measure $P(x)$) the
integral (\ref{eq:19}) yields
\begin{equation}
\label{eq:5d}
\int \d P(x) P(B_l(x))^{q-1} = \sum_{i=1}^{N} P_i P_i^{q-1}
= \chi_n(q) ,
\end{equation}
{\em From the standpoint of both data analysis and measure theory the
only significant difference between the distributions (\ref{eq:10_2})
and (\ref{eq:21}) is the lack of a partition in (\ref{eq:10_2}), and
the use of an optimal partition to define (\ref{eq:21}). Whether these
two approaches do or do not, in the limit of $l^{(n)} \rightarrow 0$
for a mathematical fractal of cardinality $t^\infty$, yield the same
generalized dimensions (whether $D_q = \nu_q$ as $l^{(n)} \rightarrow
0$) is of no importance whatsoever for the analysis of empirical
data.}

%%%%%%%%%%
\section{Lognormal distribution}
\label{sect:lognormal}

What has lognormal to do with multifractal? The question arises
because it has been asserted that the lognormal distribution is
multifractal, that it defines a spectrum of generalized dimensions
(\cite{jones:92} \& \cite{jones:origin}, \cite{jones:nonlinear}) and
an $f(\alpha)$ spectrum (\cite{frish:turbulence}).  Before answering
this question we review how and where the lognormal distribution
appears in discussions of multifractals, where a multifractal spectrum
(as defined in this paper, following Halsey \cite{halsey:86})
describes the spectrum of dimensions of nonoverlapping subsets of the
support of a probability distribution.

With the discussion of \cite{jones:92} in mind, but following
\cite{halsey:86}, let us make a largest term approximation on the
generating function (\ref{eq:5}). With $q$ fixed we first locate the
largest term in (\ref{eq:5}) by minimizing the exponent $(q \alpha
-f(\alpha))$, yielding $q=f'(\alpha(q))$. Very near (and only very
near) to the smallest exponent $\tau(q) = (q \alpha(q) -
f(\alpha(q))$, where $\alpha(q) = \tau'(q)$, we can write 
$f(\alpha) \approx \tau(q) + (\alpha -\alpha(q))^2 f''(\alpha(q))/2$.
According to a standard method (\cite{bender:advanced}) we next
replace the sum over all these nearby terms by the integral
\begin{equation}
\label{eq:21_2}
\chi_{n(q)} \approx l^{(n) \tau(q)} \int_{\delta\alpha} \d \alpha 
\rho(\alpha) \left( l^{(n)} \right)^{(\alpha - \alpha(q))^2 f''(\alpha(q))}
\end{equation}
where the range of integration $\delta \alpha$ is over the tiny region
$\delta \alpha$ in $\alpha$ containing all of the exponents $(q\alpha
- f(\alpha))$ that do not deviate from the minimum exponent $\tau(q)$
more than quadratically in $(\alpha-\alpha(q))$.  This quadratic
approximation to deviations of the exponent $(q\alpha-f(\alpha))$ from
the minimum $\tau(q)$ only works as $l^{(n)} \rightarrow 0$, in which
case the integrand in (\ref{eq:21_2}) is sharply enough peaked that,
with small error, we may extend the integration limits to plus and
minus infinity. Clearly, a locally (not globally) Gaussian
approximation to deviations from the minimum exponent $\tau(q)$ is the
same as saying that the deviations of $(N(\alpha) \left( l^{(n)}
\right)^q \alpha)$ from $\left( l^{(n)} \right)^\tau(q)$ are locally
(not globally) lognormal. This local lognormality only contributes to
the {\em prefactor} in (\ref{eq:21_2}) in the unphysical limit where
$l^{(n)} \rightarrow 0$, and not to the $f(\alpha)$ spectrum described
by the exponent $\tau(q)$. Any time that an exponent $h$ has a
Gaussian distribution then the function $p(h) = l^h$ is distributed
lognormally (see \cite{mccauley:93b} for an example from percolation
theory, where permeabilities $\kappa$ of sandstone and limestone
deposits were long thought to be approximately lognormally
distributed, with Gaussian porosities $\phi$, where $\kappa \approx
l^{\phi}$). {\em This has nothing to do with the question whether the
lognormal distribution is multifractal.}

There are two ways, {\em related mathematically to the above
approximation}, in which the lognormal distribution is called
multifractal in \cite{jones:origin} and \cite{jones:92}. Jones
\cite{jones:origin} asserts that (\ref{eq:17}) with (\ref{eq:17b}) defines
multifractal, where $\zeta_{p} = - (p-1)D_{p}$ and the $D_{p}$ are
supposed to be generalized dimensions derivable from the Halsey method
as well. {\em This constitutes an entanglement of unrelated
ideas}. Following Jones \cite{jones:origin}, we use $X=l^{h}$ in
(\ref{eq:17}) along with a lognormal distribution of $X$ to compute
$\langle l^{p h} \rangle$ (the exponent $h$ is Gaussian with mean
$\langle h \rangle$ and mean square fluctuation $\sigma^{2} = \langle
(h-\langle h \rangle )^{2} \rangle$).  Using Jone's
\cite{jones:origin} second definition of $\zeta_{p}$ (instead of his
first), where $\zeta_{p} = \ln(\langle X^{p} \rangle)/ \langle X
\rangle^{p})$, we obtain $\zeta_{p} = \exp(2 \sigma^{2} (\ln l) p
(p-1))$.  In other words a scaling law (\ref{eq:17b}) generally does
{\em not} follow: lognormal distributions per se, inserted into
(\ref{eq:17}) do not yield scale invariance (\ref{eq:17b}), because
the expected scaling exponent depends on $\ln l$. A scaling law
(\ref{eq:17b}) follows {\em only} if we restrict to lognormal
distributions where $\sigma^{2}$ is proportional to $-1/\ln l$. In
this case we obtain $D_{p} = -\langle h \rangle p$, where $D_{2}
<D_{1} < D_{0}=0$. $D_{0}=0$ is not the dimension of the support of
the lognormal distribution (where $D_H=1$), and the scaling exponents
$D_{p}$ are not the dimensions of anything else in that
distribution. (We adhere to the assumption that fractals and
multifractals are generated by deterministic dynamics, and do not
consider the so--called ``random fractals''.) Jone's refinement of
(\ref{eq:17b}) is, in this case, equivalent to the imposition of the
constraint $\zeta_{1} = 0$ in turbulence modelling (see Frisch
\cite{frish:turbulence}).

To try to model an eddy cascade in fluid turbulence one must evaluate
the average $\langle l^{p h} \rangle$ where $l$ represents the size of
an $n$th generation eddy, and $h$ is supposed to be an exponent
analogous to $\alpha$ in multifractal spectra. There, Frisch
\cite{frish:turbulence} makes a different identification than Jones,
namely, that $(p-1) D_p = \zeta_p + 3(p-1)$ for a lognormal
distribution. This yields $D_0=3$ (space-filling support) but the
exponents $D_p$ for $p \ne 0$ are not dimensions of anything in the
model. The origin of this apparent (from our standpoint) mislabeling
of scaling exponents as multifractal is that Frisch defines
$f(\alpha)$ spectra (and consequently $D_p$ spectra) differently than
we have. His definition is not designed to agree with the fractal
dimensions $f(\alpha)$ describing singular distributions ala
Halsey~\etal \cite{halsey:86}, but describes instead the {\em Cramer
function} in statistics (\cite{mandelbrot:91}). A {\em Cramer function
may exist where nothing is fractal}. A Cramer function, by
construction, describes distributions of independent random variables
$h_i$ or $\alpha_i$ the limit where $n$ goes to infinity ($l$ goes to
zero). In contrast, the indices $\alpha$ in (\ref{eq:5}) and
(\ref{eq:21}) are not random variables: they are scaling indices
describing coarsegrained probabilities (and occupation numbers $n_i =
N P_i$) $P_i=l_i^{a_i}$.

The Cramer function is a systematic way of obtaining a description
whereby $X \approx l^h$ and $P(X,l) \approx l^{-f(h)}$ via a limit
theorem in classical statistics, {\em for the case where the $h_i$ are
independent random variables}, and has no necessary connection with
the idea of spectra of Hausdorff dimensions, or generalized dimensions
derivable from spectra of Hausdorff dimensions. The Cramer function is
based on the law of large numbers and appears in classical equilibrium
statistical mechanics, for example. This approach can be used to
describe an alternative version of lognormality discussed in
\cite{jones:92}: in that case their spectrum $f(a) \approx D_0 -
(\alpha - \alpha_0)^2 / 4 (\alpha_0 - D_0)$ describes the {\em
integrand} of (\ref{eq:17}), and not a scaling law (\ref{eq:17b}) that
might follow as a consequence of actually {\em calculating} the
integral (\ref{eq:17}). A scaling law (\ref{eq:17b}) does not follow
at all without assuming arbitrarily that $\sigma^2$ varies as $-1/\ln
l$.  With that restriction we obtain $(p-1)D_p = \zeta_p - (p-1) D_0$
with $\zeta_p = - \alpha_0 p$, analogous to Frisch's result quoted
above, but where we should now choose $D_0=1$ in order to describe the
support of the lognormal distribution (otherwise, $D_0$ is not the
fractal dimension of anything in the model). Even with this choice the
remaining $D_p$ are not fractal dimensions, even for $p =1,2$, of any
aspect of the lognormal distribution. However, the {\em same}
lognormal distribution can be understood from an entirely {\em
different} standpoint: namely, as a combination of the Gaussian
integrand with the exponent $\tau(0) = - D_0$ in equation
(\ref{eq:21_2}) of Halsey et al above. In this case $D_0$ is is {\em
not} the Hausdorff dimension of the support of the lognormal
distribution, and the linear spectrum $D_p$ describes {\em only} the
region near the peak of an unknown $f(\alpha)$ spectrum, a spectrum of
box--counting dimensions where $f_{\rm max} = D_0$ (from the
standpoint of eqn. (\ref{eq:21_2}) above one would interpret the
$\zeta_p$ spectrum derived from the lognormal distribution in
turbulence modelling as the fragment of an $f(\alpha)$ spectrum that
is valid only for very small values of $p$, near $p=0$, although this
is certainly {\em not} the traditional interpretation). On the other
hand, however, the {\em generalization} of the lognormal approximation
represented by Jone's \cite{jones:nonlinear} equations (14) and (15),
and (30), are not definitions of $f(\alpha)$ ala Halsey~\etal
\cite{halsey:86}, but represent instead the idea of a Cramer function
in classical statistics (see Frisch '95): {\em unless $f(\alpha)$
arises as the spectrum of Hausdorff dimensions from an infimum
condition on partitions, then $f(\alpha)$ is not (by the Halsey
\cite{halsey:86} definition) a multifractal spectrum.} This entanglement 
of different ideas did not originate in cosmology: one
of the authors of Halsey~\etal \cite{halsey:86} later used the Cramer
function, called it ``multifractal'', and cited Halsey~\etal
\cite{halsey:86} as the reference (\cite{kadanoff:89}).

One can choose to follow Halsey~\etal \cite{halsey:86} in defining
$f(\alpha)$ and $D_q$, or one can follow Mandelbrot
\cite{mandelbrot:91} in defining ``$f(\alpha)$ and $D_q$'' via Cramer
functions \cite{frish:turbulence}, but one should not mix these two
different definitions together without comment as is done in
\cite{jones:origin}. We recommend the definition of multifractal given
in this paper as the standard because, in that case, $f(\alpha)$ is
always the Hausdorff dimension of a subset of the support of the
empirical distribution $P(x)$. The necessity of an optimal partition
in order to define $f(\alpha)$ is implicit in Halsey~\etal
\cite{halsey:86} (see their ``infimum'' requirement) but was not
emphasized strongly enough at that time. The role played by the
infimum requirement became clear only after the later work on generating
partitions in nonlinear dynamics (\cite{feigenbaum:88},
\cite{cvitanovic:88} and \cite{artuso:90}), which is little--known 
within the community of cosmologists.

When is a distribution $P(x)$ multifractal? To answer this question
consider any distribution $P(x)$, empirical or theoretical, where
$P(0)=1/N$, $P(1)=1$, and $P(x)$ is nondecreasing. For idealized
differentiable distributions we have $N=2^\infty$ (which is the same
as $10^\infty$, etc.) and $P(0)=0$. $P(x)$ need not be differentiable,
however, and generally isn't. In order to determine whether $P(x)$
``is multifractal'' (admits a decomposition of its support into
interwoven fractals with different dimensions $f(\alpha)$) one must
determine the optimal partition and form the difference $\Delta P(x)$
over each interval in that support to obtain the hierarchy of
histograms $\{ P_i \}$ (for a differentiable distribution any
space-filling partition will do the job). Having done that, one then
investigates whether $P_i \approx l_i^{\alpha_i}$ holds over
$N(\alpha) \approx l^{-f(\alpha)}$ intervals in the support as the
interval sizes are systematically reduced. All fractal distributions
(for points on a line) have an density $\rho_i \approx l_i^{\alpha_i
-1 }$ that is dense with singularities because $\alpha_i < 1$, and if
the distribution is multifractal then the indices $\alpha_i$ will vary
over the support according to $N(\alpha) \approx l^{-f(\alpha)}$. A
highly fragmented (spiky) coarsegrained density is typical of a
multifractal distribution $P(x)$.

Gaussian distributions are not multifractal. Neither are lognormal
distributions. No smooth, differentiable distribution is multifractal
because, by definition, such a distribution has a smooth density on a
support with integer dimension $D_0$. Smooth distributions can be
differentiated everywhere, corresponding to the requirement that
$f(\alpha) = \alpha = D_0$ holds everywhere. The Cantor function
$P(x)$ in part \ref{sect:empirical_dist} describes clustering and
voids), the binomial distribution with $p_1 \ne p_2$ on a
space--filling support describes intermittence without voids (see
\cite{meneveau:87} for a physical example), but the lognormal
distribution cannot describe voids because it is differentiable. The
distribution defined by the density $\rho(x) = \d P(x) /\d x = \left(
x(1-x) \right)^{-1/2}$ is not differentiable at $x=0$ and 1 and is
bifractal (Halsey~\etal
\cite{halsey:86}): $\alpha= f(\alpha) =1$ for $0 < x < 1$ but
$\alpha=1/2$, $f(\alpha) = 0$ for $x=0$ and 1. The bifractal
staircase of She \etal \cite{she:92} shows coalescence without voids
only because a continuum of initial conditions was used rather than a
finite number (blocks of initial conditions with voids should also
produce coalescence with voids in that model).

Summarizing, nontrivial $f(\alpha)$ spectra guarantee an intermittent
probability density, one corresponding to a nondifferentiable
probability distribution $P(x)$. All fractal distributions have
singular densities. Having a fractal support guarantees a singular
density $\rho \approx l^{\alpha-1}$ with $0< \alpha < 1$, but
multifractal scaling can also hold for inhomogeneous distributions
(like those having statistical independence with uneven probabilities)
on a space-filling support.

%%%%%%%%%%
\section{Homogeneity, coarsegraining and hydrodynamics}
\label{sect:homogeneity}
The singular ``pointwise'' density of matter $\rho(x,y,z)$ in any
epoch is determined by the empirical staircase distribution
$P(x,y,z)$, the generalization of (\ref{eq:3b}) to three dimensions. A
necessary but {\em insufficient} condition for a coarsegrained density
that is smooth enough to be approximable by a differentiable function
is that the support of the empirical distribution is space--filling,
meaning $D_H =3$.

No information about $D_H$ is provided by the correlation integral
exponent $\nu$ (or by the correlation dimension $D_2$) unless, (1) the
support is monofractal and the distribution is uniform ($\alpha =
f(\alpha) = D_H < 3$), or (2) the support is space--filling and the
distribution is uniform or at least differentiable ($\alpha =
f(\alpha) = D_H = 3$). In both cases $\nu = D_2 = D_H$. Otherwise, we
know only that $\nu < D_H$ and $D_2 < D_H$. An increase in $\nu$ with
an increase in scale from intermediate toward cosmologic scales,
$l^{(n)} \gg 1000 \hMpc$, even if it would be found in the data, would
not tell us anything about $D_H$ unless we would find that $\nu=3$. So
long as $\nu < 3$ no information about $D_H$ is provided by $\nu$. An
increase of $\nu$ with increasing scale (as is claimed in
\cite{martinez:multiscaling-proc}, \cite{borgani:cluster-scaling}) would
not imply that there exists a large scale coarsegraining where $D_H
=3$. If $D_H < 3$ then the approximation of the empirical distribution
by a differentiable one is impossible.  The same conclusion follows if
$\nu < D_H =3$.

Fractal scaling of galaxies in an intermediate range $1 \le r \le 1000
\hMpc$, e.g. would still allow for the possibility of nonfractal
matter distributions at the largest (or ``cosmologic'') scales. An
empirical matter distribution $P(x,y,z)$ that shows no clusters and
voids ``at large enough scales'' $l^{(n)}$ would require a support
with dimension $D_H = 3$. Whether galaxies are distributed more or
less uniformly over a nonuniform space-filling support $\{ l_{x,i}
l_{y,i} l_{z,i}\}$ is then a question of whether $P_i \approx l_{x,i}
l_{y,i} l_{z,i}$ holds over enough generations of the hypothetical
``cosmologic-scale support'' so that one can define the derivatives of
densities normally used in hydrodynamics. Only for a uniform support
would this condition reduce to the requirement of statistical
independence with equal probabilities, $P_i \approx N_n^{-1}$. This is
the requirement for large-scale uniformity ($\rho(x,y,z) \approx
\mbox{constant}$) stated in the language of dynamical systems theory.

A necessary condition for homogeneity in a given direction $x$, at
cosmological scales $l^{(n)} \gg 1000 \hMpc$, can also be stated as
follows: on what scale $l^{(n)}$ of cosmologic coarsegraining can a
staircase $P(x)$ of $N$ steps ($P(x)$ denotes the empirical
distribution $P(x,y,z)$ with $y$ and $z$ held constant) be
approximated by a differentiable distribution, $P'(x) \approx
\rho(x)$, where $\rho(x)$ is smooth, approximately analytic? There are
two requirements: the number $n_i$ of data points in each interval
must be very large, and the spacing between points cannot be very
different from $\Delta x \approx 1/N$. This is the same as saying that
the steps in $P(x)$ are nearly uniform and of very small width, and
lie approximately on the straight line $P(x) \approx x$ with slope
$\rho(x) \approx 1$ (uniform density). If the pointwise spacing is not
exactly $\Delta x \approx 1/N$, but $D_H =3$ and there are no voids
and clustering, then the staircase may be a smooth deformation of the
constant density distribution $P(x) \approx x$ with variable but
nearly smooth slope $\rho(x) \approx P'(x)$ approximating a nonuniform
smooth density.
The necessary and sufficient condition for large--scale homogeneity,
stated in the language of statisticians is given in Stoyan \etal
\cite{stoyan:stochgeom}. Contrary to the advice of part 9 in 
\cite{coleman:fractal} we point out that ``pencil beam surveys'' 
generally cannot be treated as one--dimensional cuts. In order to
qualify as a one dimensional cut, the maximum width of a pencil beam
survey should be on the order of the size of a galaxy.

If we think in terms of hydrodynamic models of clustering, then
space--filling supports, by Liouville's theorem, require conservative
dynamical systems. A smooth density at large scales cannot be
the result of dissipative hydrodynamics.

No empirical test can be performed globally on the scale of the
universe. The best that one can hope for is to gain information about
the different local distributions of matter for various different
samples of galaxies and clusters of galaxies at scales $r \gg 1000
\hMpc$ and test them for homogeneity and isotropy (our
Euclidean language is applicable locally in a curved space-time). If
the scales required to exhibit evidence for homogeneity and isotropy
should fall beyond the inherent limitations on all future
observations, then the cosmological principle is not falsifiable and
is only a matter of belief.

A more useful question is how to combine fractal or multifractal
distributions with hydrodynamics, now defining the density
\begin{equation}
\label{eq:mass_density}
\rho(\vec{x}) = \sum_{i=1}^N m_i \delta(\vec{x}-\vec{x_i})
\end{equation}
to include the masses of galaxies, as is done in part 6 of
\cite{coleman:fractal}. In a related publication \cite{pietronero:96}
it is asserted that 
``\dots phenomena in which {\em intrinsic self--similar irregularities
develop at all scales} and fluctuations cannot be described in terms
of analytic functions. The theoretical methods used to describe this
situation could not be based on ordinary differential equations
because self--similarity implies the absence of analyticity and the
familiar mathematical physics becomes inapplicable.''
These claims are patently false: There is certainly no reason why one
cannot study the dynamics of eq.~(\ref{eq:mass_density}) as an N--body
problem using nonlinear differential equations, as was done by
mathematicians from Laplace to Poincar\'{e} and beyond! Why was such a
sweeping statement made to begin with? Clearly, by the extrapolation
length scales to zero, to the empirically and
physically meaningless mathematical limit where, essentially, two
galaxies occupy the same position. Only in this limit are fractal
densities nonanalytic enough to be completely nondifferentiable.

In fact the coarsegrained picture of empirical distributions
formulated in parts \ref{sect:empirical_dist} and
\ref{sect:multi_empirical} above leads in principle to a hydrodynamic
description in terms of the usual differential equations of
mathematical physics. At any desired resolution $\approx l_n$, simply
represent the density by
\begin{equation}
\label{eq:mass_density_coarsegrained}
\rho(\vec{x}) = \sum_{i=1}^{N_n} \rho_i \chi_{l_i}(\vec{x}-\vec{x_i})
\end{equation}
where (see eq.~(\ref{eq:mass_density}) above) $\rho_i$ is the
coarsegrained mass density for a partitioning $\{ l_i \}$. One can
certainly study the stability of this distribution (taken as an
initial condition) via the usual differential equations of
hydrodynamics, even if this may require solutions in the weak (or
distribution) sense. The results will not be correct at length scales
smaller than the scales $\{ l_i \}$, but at smaller scales (down to
$l_{\rm min} > 0$) one can increase $N_n$ (decreasing the size of
intervals $l_i$) and again study stability questions via a finer
grained density of the form of (\ref{eq:mass_density_coarsegrained}).
Uniform densities on space--filling supports in Newtonian cosmology
are unstable if the universe is open, stable if the Euclidian manifold
is a flat 3--torus \cite{buchert:averaging}. In other words,
homogeneity is globally unstable in an open Newtonian universe.

%%%%%
\section{Platonic expectations?}

The standard model of cosmology (\cite{weinberg:gravitation},
\cite{peebles:principles}) is a paradigm of simplicity: the simplest
possible solution of Einstein's field equations (global integrability
based upon global symmetry) is combined with what Feynman has called
``the usual initial conditions of physics'': random or thermal initial
conditions. Feynman pointed out that biologists, geologists, and
astronomers know that the usual initial conditions of physics (and
integrable dynamics, one must add) cannot be used to explain most of
the phenomena that are observed in nature (\cite{feynman:gravity}).
Cosmologically seen, far from equilibrium phenomena occur at
relatively small scales. That these nonequilibrium nonuniformities at
small scales should be consistent with perfect symmetry and random
initial conditions at the largest scales is not at all clear. If it
would be true then, as Plato \cite{plato:timaeus} believed, the
heavens would be perfect while all disorder is confined to the
``sphere of the earth'' (extended a bit, out to 150~\hMpc, at least).

It is not necessary to assume that the galaxy distribution requires
the nineteenth century notion of randomness except perhaps as a
sometimes convenient approximation to deterministic chaotic
dynamics. We now understand how even the simplest chaotic dynamical
systems can generate all possible historgams that can be constructed
empirically (\cite{mccauley:chaos}). By randomness we mean a breakdown
of the space--time description of cause and effect (as in quantum
mechanics). Statistical independence (\cite{kac:}), in contrast, is a
separate idea that occurs in deterministic dynamics. Physics in the
last twenty years has begun to follow more the path laid out by
Poincar\'{e} (\cite{mccauley:classical}), deviating from the
traditional path set down by Boltzmann (contrast the traditional
emphasis on randomness found in \cite{mandelbrot:fractal} and in
\cite{zeldovich:almighty}, for example, with the perspective on
randomness expressed in \cite{feigenbaum:80}). 

A deterministic dynamical system (like the differential equations that
generate the characteristic curves of Newtonian or Einsteinian
cosmology) can generate inhomogeneity that need not be fractal. A
dynamical system far from thermodynamic equilibrium does not generate
a unique probability distribution, but instead generates infinitely
many different classes of distributions depending on classes of
initial conditions. We would have no way to discover ``the initial
conditions of the universe'' other than by accurate backward
integrations in time starting from the present (empirically unknown)
distribution of matter via the correct equations of motion. The laws
of physics alone do not tell us anything about initial conditions
(\cite{wigner:symmetries}). In the far from equilibrium case, on small
scales, we know that mother nature has not chosen ``the usual initial
conditions of physics'' (trees don't grow from thermal equilibrium
initial conditions, but arise instead from strong driving combined
with dissipation).

Even if we knew the present cosmologic--scale distribution of matter
the question whether the global matter distribution could have been
generated from a thermally equilibrated initial state cannot be
answered by N--body simulations, because no existing computer can
reproduce, in backward integration in time, even the first digits of
those initial conditions after integrations forward in time over
billions of years. Accurate backward integrations were actually
accomplished, for unknown reasons and for a very restricted range of
energies, by building a special computer to try to simulate the
evolution of the solar system via a chaotic symplectic map over
millions of years (\cite{sussmant:92}). One object was to try to
understand the initial conditions that initially fascinated
Kepler\footnote{Kepler was denied research support for his Platonic
project to explain the initial conditions of the solar system on the
basis of geometry alone (\cite{caspar:kepler}).},
the so-called Titius ``law'' (see also \cite{weizsaecker:44}).

The characteristic curves of the partial differential equations of
Newtonian cosmology are generated by a dynamical system with a phase
space of at least six dimensions (three degrees of freedom).  The
Lagrangian method \cite{buchert:averaging} studies characteristics via
backward integration in time, using the fact that initial conditions
are trivially conserved along streamlines.  Chaotic dynamics requires
only a three dimensional phase space.  Complex dynamics, dynamics
equivalent to a universal computer (no scaling laws, no generating
partition, nothing to aid in forecasting the future statistically) may
occur in certain Newtonian systems with only three degrees of freedom
(\cite{moore:90}). There, scaling laws, attractors, and the
cardinality of strange sets can at best be defined only locally, if at
all.
It is unknown whether (any form of) fluid turbulence or the Newtonian
three body problem fall into the complexity category. Expecting the
universe to be describable by a completely integrable dynamical
system, even at the largest scale of coarsegraining, seems unlikely in
the light of what we now understand about deterministic dynamics.  We
are reminded that the cosmological principle is not demanded by any
known laws of physics, and is not itself a separate law of nature.

This article was written with the benefit of hindsight. The author
belongs to the subset of physicists who, in the past, has claimed
evidence for $\tau(q)$ on the basis of log--log plots that did satisfy
the Geilo Criterion.

%%%%%%%%%
\section*{Acknowledgement}
I am very grateful to Martin Kerscher for generating the data analysis
in Section \ref{sect:data}, to Herbert Wagner and Thomas Buchert for
criticism and discussions and to Herbert Wagner and Universit\"at
M\"unchen for guestfriendship during Sebt. -- Feb. of my 1996--97
sabbatical year where I had the chance to learn some cosmology. The
focus necessary to carry this project through was generated by the
Oct.~1996 Schlo\ss Ringberg conference on cosmology. Thomas Buchert
made it possible for me to attend that conference, while the
encouragement to enter into the debate over fractals in galaxy
statistics came from Herbert Wagner.
We are grateful to Bernard Jones for some comments, and to Francesco
Sylos--Labini and Luciano Pietronero for an exchange of ideas via
email.

%%%%%%%%%%%%%%%%
%\bibliographystyle{scaling}
%\bibliography{scaling}

\ifx\undefined\bysame
\newcommand{\bysame}{\leavevmode\hbox to3em{\hrulefill}\,}
\fi

\end{document}